\documentclass[journal]{IEEEtran}
\usepackage{amsmath,amsfonts}
\usepackage{algorithmic}
\usepackage{algorithm}
\usepackage{array}
\usepackage[caption=false,font=normalsize,labelfont=sf,textfont=sf]{subfig}
\usepackage{textcomp}
\usepackage{stfloats}
\usepackage{url}
\usepackage{verbatim}
\usepackage{graphicx}
\usepackage{cite}
\usepackage{booktabs}
\usepackage{epstopdf}
\DeclareMathOperator*{\argmax}{argmax}

\begin{document}

\title{Joint Design of Access and Backhaul in Densely Deployed MmWave Small Cells}

\author{Ziqi~Guo,
        Yong~Niu,~\IEEEmembership{Senior~Member,~IEEE,}
        Shiwen~Mao,~\IEEEmembership{Fellow,~IEEE,}
        Ruisi~He,~\IEEEmembership{Senior~Member,~IEEE,}
        Ning~Wang,~\IEEEmembership{Member,~IEEE,}
        Zhangdui~Zhong,~\IEEEmembership{Fellow,~IEEE,}
        and~Bo~Ai,~\IEEEmembership{Fellow,~IEEE}
\thanks{Copyright (c) 2015 IEEE. Personal use of this material is permitted. However, permission to use this material for any other purposes must be obtained from the IEEE by sending a request to pubs-permissions@ieee.org. This work was supported in part by the Fundamental Research Funds for the Central Universities under Grant 2022JBXT001 and Grant 2022JBQY004; in part by the National Key Research and Development Program of China under Grant 2020YFB1806903; in part by the National Key Research and Development Program of China under Grant 2021YFB2900301; in part by the National Natural Science Foundation of China under Grant 62221001, Grant 62231009, Grant U21A20445 and Grant 61771431; in part by the Fundamental Research Funds for the Central Universities 2023JBMC030. (\emph{Corresponding author: Y. Niu, B. Ai.})}
\thanks{Ziqi Guo is with the State Key Laboratory of Advanced Rail Autonomous Operation, Beijing Jiaotong University, Beijing 100044, China, and also with the Collaborative Innovation Center of Railway Traffic Safety, Beijing Jiaotong University, Beijing 100044, China (e-mail: 21120053@bjtu.edu.cn).}
\thanks{Yong Niu is with the State Key Laboratory of Advanced Rail Autonomous Operation, Beijing Jiaotong University, Beijing 100044, China (e-mail: niuy11@163.com).}
\thanks{Shiwen Mao is with the Department
of Electrical and Computer Engineering, Auburn University,
Auburn, AL, 36849-5201 USA (e-mail: smao@ieee.org).}
\thanks{Ruisi He, Zhangdui Zhong, and Bo Ai are with the State Key Laboratory of Advanced Rail Autonomous Operation, Beijing Jiaotong University, Beijing 100044, China, and also with the Beijing Engineering Research Center of High-speed Railway Broadband Mobile Communications, Beijing Jiaotong University, Beijing 100044, China (e-mails: ruisi.he@bjtu.edu.cn; zhdzhong@bjtu.edu.cn; aibo@ieee.org).}
\thanks{Ning Wang is with the School of Information Engineering, Zhengzhou University, Zhengzhou 450001, China (e-mail: ienwang@zzu.edu.cn).}}

\maketitle

\begin{abstract}
With the rapid growth of mobile data traffic, the shortage of radio spectrum resource has become increasingly prominent. Millimeter wave (mmWave) small cells can be densely deployed in macro cells to improve network capacity and spectrum utilization. Such a network architecture is referred to as mmWave heterogeneous cellular
networks (HetNets). Compared with the traditional wired backhaul, The integrated access and backhaul (IAB) architecture with wireless backhaul is more flexible and cost-effective for mmWave HetNets. However, the imbalance of throughput between the access and backhaul links will constrain the total system throughput.
Consequently, it is necessary to jointly design of radio access and backhaul link. In this paper, we study the joint optimization of user association and backhaul resource allocation in mmWave HetNets, where different mmWave bands are adopted by the access and backhaul links. Considering the non-convex and combinatorial characteristics of the optimization problem and the dynamic nature of the mmWave link, we propose a multi-agent deep reinforcement learning (MADRL) based scheme to maximize the long-term total link throughput of the network. The simulation results show that the scheme can not only adjust user association and backhaul resource allocation strategy according to the dynamics in the access link state, but also effectively improve the link throughput under different system configurations.
\end{abstract}


\section{Introduction}\label{S1}

\IEEEPARstart{I}{n} recent years, mobile data traffic has grown rapidly, which leads to escalating demands for wireless spectrum. Heterogeneous cellular networks (HetNets) provide a promising solution to improve network capacity and meet the traffic demand~\cite{01}. By deploying small cells underlaying macro cell, user equipment (UE) can be associated with a closer base station (BS) to obtain better quality of service (QoS). However, there is a compelling need to provide enough bandwidth to meet the growing user demand in HetNets. A promising approach is spectrum expansion, i.e., exploiting the millimeter wave (mmWave) band from 30 GHz to 300 GHz~\cite{02,03}. The mmWave band offers huge amount of bandwidth, which can be utilized to support bandwidth-intensive applications. However, due to the high carrier frequency, mmWave communications suffer from more severe propagation loss than sub-6 GHz systems. Thus, both the transmitter and receiver should adopt high gain directional antennas to achieve directional transmission. Besides, owing to the small wavelength and the directional transmission, mmWave links are sensitive to the random blockage by the obstructions in the environment.

In HetNets with small cells densely deployed, deploying wired fiber backhaul is costly~\cite{04}. Therefore, 3GPP advances the integrated access and backhaul (IAB) architecture for 5G cellular network~\cite{05}, which provides wireless backhaul and supports wireless access and backhaul simultaneously with shared radio resources. Two scenarios have been considered for IAB by 3GPP: the in-band backhaul scenario and the out-of-band backhaul scenario~\cite{06}. In a mmWave system, the in-band backhaul scenario means that access and backhaul share the same mmWave band to improve spectrum utilization and achieve closer integration. However, this resource sharing makes the available spectrum limited, and the interference between access and backhaul is introduced into the system, which is more serious in small cells densely deployed~\cite{07b,08}. In comparison, an out-of-band backhaul scenario with two different mmWave bands allocated for access and backhaul will alleviate the resource pressure resulting from the dense deployment of small cells and eliminate the backhaul-access link interference. Consequently, the network throughput can be further improved, and the design and optimization of HetNets can be more flexible. Besides, considering the different characteristics of different mmWave bands, we can choose the appropriate mmWave band for access and backhaul respectively according to the actual propagation environment, which exploits the value of mmWave to a greater extent~\cite{08b}. Therefore, this scenario has great potential in future networks with higher data rate requirements and denser deployments.

For the mmWave HetNets, the performance of the system is influenced by both the access link and the backhaul link, so it is important to jointly design both of them~\cite{09}. Specifically, there are two vital problems to be solved. The first one is the backhaul resource allocation. Because of the randomness and dynamics of wireless networks, the design of resource allocation strategy needs to be robust to deliver satisfactory performance with regard to the network throughput and resource utilization. The other problem is the user association. With a properly designed user association strategy, users can compare the QoS of different BSs and choose to associate with the best choice, so as to improve the user experience. When a BS serves too many users, some of its users can be handed over to an adjacent available BS to reduce the traffic load. However, in the mmWave HetNets, the user association and the backhaul resource allocation are coupled. To this end, how to jointly optimize backhaul resource allocation and user association in the HetNets is a key problem~\cite{10,11,12,13}.

With the development of artificial intelligence, there is an increasing interest in applying learning based algorithms to address wireless communications problems~\cite{13b,14,15,16,17}. One of such effective techniques is reinforcement learning (RL)~\cite{20}. By interacting with environment and training agents, RL can evaluate policies and adaptively select the optimal policy. The RL based algorithm does not need to know detailed information of the environment, and is adaptive to the changes in the environment. The classical RL, such as Q-learning, has been shown to achieve superior performance in small-sized systems~\cite{21}. However, as Q-learning needs to store the Q-value of each state-action pair in a Q-table, such storage and computing cost could be prohibitively high for systems with high-dimensional state and action spaces. To address this problem, a deep neural network (DNN) can be leveraged to approximate the Q-value, thus the Q-table can be replaced. Such a model is usually referred to as deep reinforcement learning (DRL)~\cite{22}. Moreover, when there are more than one agents in the system, the state of environment depends on the joint actions of all the agents~\cite{23}. Such a model, termed multi-agent reinforcement learning (MARL), has a great potential in solving distributed optimization problems such as resource allocation and user association~\cite{26}.

In this paper, we conduct research on the joint design of access and backhaul for the mmWave HetNets. The access links and the backhaul links adopt different mmWave bands in the two-layer HetNets, which can be considered as out-of-band IAB networks with single-hop backhaul. Small cells are densely deployed in the HetNets and
the access links are considered to be affected by the random blockage. We focus on the joint optimization of the user association and the backhaul resource allocation to improve long-term total link throughput in the HetNets. We develop an effective MADRL based method, which allows each UE to select the associated small base station (SBS) and determine the backhaul resource requirements based on its state observations. The method does not require the full channel state information (CSI) and uses a distributed architecture to improve training efficiency. The contributions of this paper are summarized as follows:

\begin{itemize}
  \item We focus on the joint optimization of backhaul resource allocation and user association in a HetNet with the dense deployment of mmWave small cells. The access links and the backhaul links adopt two different mmWave bands and the access links are subject to random blockage. We formulate the joint optimization as a mixed integer nonlinear programming (MINLP) problem, to maximize the long-term total link throughput in the HetNet.

  \item We develop a MADRL-based method to solve this joint optimization problem. Specifically, we consider each UE as an agent and define the state, action, and reward for UEs. Then, with the help of the double deep Q-learning algorithm (DDQN), each UE learns an effective joint optimization strategy to maximize the long-term total link throughput. Through distributed training, each UE can make decisions independently based on partial observations of the environment and adjust the joint optimization strategy in time when the link state changes.

  \item We evaluate the MADRL scheme and show that the proposed learning framework has good convergence performance. When the link state changes, the network throughput performance can be guaranteed by adjusting the joint optimization strategy. Besides, the scheme contributes to a better balance between access throughput and backhaul throughput, and also achieves higher total link throughput in various network scenarios with different numbers of UEs and SBSs over several baseline schemes.

\end{itemize}

The rest of the paper is organized as follows. Section~\ref{S2} reviews related work. The system model and problem formulation are presented in Section~\ref{S3}. We present the joint design scheme based on MADRL for both radio access and backhaul network in Section~\ref{S4}. Our simulation results are discussed in Section~\ref{S5}. Section~\ref{S6} concludes this paper. We list the abbreviations commonly appeared in this paper in Table~\ref{Abbreviations}.

\begin{table}[!t]
\caption{List of Abbreviations}
\vspace{-3mm}
\label{Abbreviations}
\centering
\begin{tabular}{cm{6cm}}
\toprule 
Notation & Description\\
\midrule 
BS & base station\\
DA & distance based user association and equal backhaul bandwidth allocation\\
DDQN & double deep Q-learning\\
DNN & deep neural network\\
DQN & deep Q-learning\\
DRL & deep reinforcement learning\\
HL & heuristic based user association and load based backhaul bandwidth allocation\\
IAB & integrated access and backhaul\\
LOS & line-of-sight\\
MADDQN & multi-agent double deep Q-learning\\
MADRL & multi-agent deep reinforcement learning\\
MARL & multi-agent reinforcement learning\\
MBS & macro base station\\
MDP &  Markov decision process\\
MINLP & mixed integer nonlinear programming\\
NLOS & non-line-of-sight\\
ReLU & rectified linear unit\\
RL & reinforcement learning\\
SADRL & single-agent deep reinforcement learning\\
SBS & small base station\\
SINR & signal-to-interference-plus-noise ratio\\
SL & SNR based user association and load based backhaul bandwidth allocation\\
SNR & signal-to-noise ratio\\
TDMA & time division multiple access\\
UE & user equipment\\
\bottomrule 
\end{tabular}
\vspace{-6mm}
\end{table}

\section{Related Work}\label{S2}

There have been several related works studying joint resource allocation and user association for HetNets.
For example,
Fooladivanda \emph{et al.}~\cite{27} investigated the performance of different user association policies under three predefined spectrum allocation strategies.
Lin \emph{et al.}~\cite{28} focused on jointly optimizing user association and spectrum allocation in both downlink and uplink to maximize network utility.
Chen \emph{et al.}~\cite{29} aimed to maximize the system sum rate and designed a distributed algorithm for the joint optimization.
Zhuang \emph{et al.}~\cite{30} proposed an optimization-based framework to reduce network energy consumption.
With huge available bandwidth, mmWave can be adopted in HetNets, significantly increases the network capacity. In~\cite{31} and~\cite{32}, the user association and the resource allocation were jointly optimized in the HetNets with the coexistence of sub-6 GHz BSs and mmWave BSs.
Liu \emph{et al.}~\cite{08b} investigated the joint user association and resource allocation in mmWave HetNets under two access modes: single-band access and multi-band access.

However, the above works only considered the optimization of the access side. In HetNets with wireless backhaul, the balance of the access and backhaul link throughput is essential for higher total throughput. Therefore, it is necessary to consider a joint design of access and backhaul. Many works have considered the scenarios where access and backhaul operate at the same frequency band.
Liu \emph{et al.}~\cite{10} proposed an iterative algorithm for the joint optimization of user association and resource allocation in in-band full-duplex HetNets.
Khodmi \emph{et al.}~\cite{11} adopted non-cooperative game theory to solve the joint power allocation and user association problems, aiming to guarantee throughput balance between the access and backhaul links.
Su \emph{et al.}~\cite{12} adopted a distributed optimization algorithm based on primal and dual decomposition to jointly optimize the user association
and the backhaul bandwidth allocation.
Liu \emph{et al.}~\cite{13} proposed a coalition game based joint user association and bandwidth allocation algorithm to maximize network sum rate.
However, the fact that access and backhaul share the same frequency band makes frequency resources limited and introduces backhaul-access interference. The interference is exacerbated by the dense deployment of small cells, which limits the improvement of throughput~\cite{07b}.
There have been some works focused on the HetNets with mmWave backhaul and sub-6 GHz access~\cite{34,35,36}. Despite the elimination of the backhaul-access interference, the limited bandwidth of the sub-6 GHz band makes it difficult to support higher data rate transmissions. In addition, allocating different mmWave bands to access and backhaul may be another effective solution. However, few works have focused on the joint optimization of access and backhaul under this solution.

Besides, model-free RL has been widely used to solve optimization problems in wireless communication~\cite{26}.
Feng \emph{et al.}~\cite{37} applied deep Q-learning (DQN) to find the resource allocation strategy under different system states.
Wei \emph{et al.}~\cite{38} proposed a actor-critic RL algorithm to obtain the user scheduling and resource allocation scheme under the consideration of the stochastic property of wireless channel conditions and renewable energy arrival rates.
Shen \emph{et al.}~\cite{17} established a DRL framework to optimize the resource allocation and scheduling for time-sensitive traffic in a 5G system subject to mmWave channel variations.
However, these works considered centralized optimization, which induces a heavy computational pressure on the central controller. Besides, the communication overhead of collecting global information cannot be ignored. In comparison, distributed optimization based on MARL are more advantageous in the large-scale dynamic HetNets. For example,
Zhao \emph{et al.}~\cite{39} regarded each user as an agent and proposed a distributed optimization method based on MADRL to achieve the jointly optimal resource allocation and user association strategies in HetNets.
Yang \emph{et al.}~\cite{40} proposed a multi-agent dueling DQN-based algorithm combined with distributed coordinated learning to jointly optimize device association, spectrum allocation, and power allocation in dynamic HetNets.
Sana \emph{et al.}~\cite{41} developed a MADRL based user association scheme under the time-varying nature of the mmWave channels for dense HetNets. However, these works did not consider the joint design of access and backhaul.

Motivated by these prior works, we focus on the joint design of access and backhaul in densely deployed small cells, where different mmWave bands are used for access and backhaul. We propose a MADRL-based scheme for joint user association and backhaul bandwidth allocation over the access and backhaul networks. The proposed scheme considers the dynamic characteristics of mmWave communications as well as the interaction between backhaul bandwidth allocation and user association in HetNets, aiming
to maximize the long-term overall system throughput.

\section{System Model and Problem Formulation}\label{S3}

\subsection{System Model}\label{S3-1}

\begin{figure}[htbp]
\begin{center}
\includegraphics*[width=0.9\columnwidth,height=2.0in]{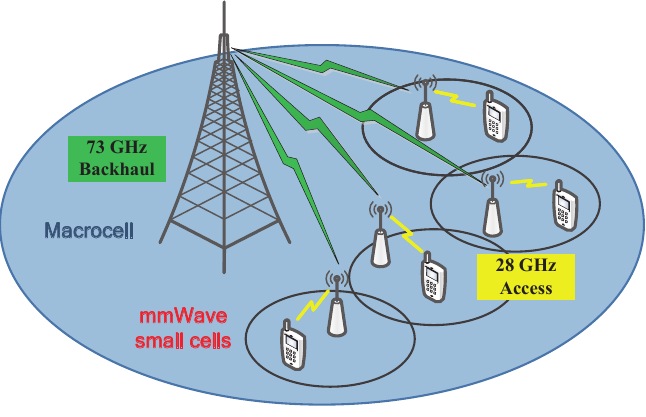}
\end{center}
\vspace{-2mm}
\caption{Dense deployment of small cells underlying the macrocell network with different mmWave bands for access and backhaul.}
\vspace{-2mm}
\label{fig:1}
\end{figure}

Consider a two-tier downlink HetNet composed of a macro base station (MBS), $S$ SBSs, and $N$ randomly located UEs, as shown in Fig.~\ref{fig:1}. The MBS is connected to the core network through optical fiber. The SBSs are densely deployed within the coverage of the MBS. Data transmission is carried out between the MBS and SBSs through mmWave backhaul links. We assume that the MBS and SBSs can be appropriately deployed to avoid the blockage of the backhaul link~\cite{09}. Thus, the stable line-of-sight (LOS) connection can be established between the MBS and each SBS without relaying through other SBSs. Each UE can be associated with a SBS and served by the associated SBS through a mmWave access link\footnote{
It is worth noting that we do not consider the direct communications between the MBS and UEs. Although this is beneficial to improve the received signal strength for UEs near the MBS, few UEs can benefit from such an improvement due to the large coverage area of the macro cell and the severe path loss and signal blockage suffered by the mmWave signal. In addition, the introduction of direct communications causes more severe interference to UEs served by SBSs, thus degrading their throughput performance.}.

In this paper, we adopt the 28 GHz band and the 73 GHz band for the data transmission of access and backhaul, respectively. Both the BSs and UEs are assumed
to be equipped with antenna arrays for performing directivity beamforming. Besides, since mmWave is sensitive to the blockage of environmental obstructions, we further assume that the BSs and UEs are also equipped with omnidirectional antennas in sub-6 GHz for reliable transmission of the transmission requests and signaling information~\cite{42}. Let $U \!\!=\!\! \{1, 2, ..., N\}$ and $B \!\!=\! \{1, 2, ..., S\}$ denote the sets of UEs and SBSs, respectively.

\vspace{-4mm}

\subsubsection{Access Transmisson Model}
The access transmissions are performed in 28 GHz. In the access network, frequency resources are multiplexed in different small cells. Time is assumed to be partitioned into multiple superframes and each superframe consists of many nonoverlapping time slots. Each SBS uses time division multiple access (TDMA) to serve its associated UEs, which means it serve one UE in each time slots. Moreover, each SBS evenly allocates time slots to its associated UEs in the small cell. Suppose that at each time $t$
\footnote{Note that the time $t$ corresponds to the time when the link blockage state changes. The change is a large-scale characteristic of the wireless channel. The time scale of a time $t$ is much larger than the superframe length of the access link.}, each SBS can be associated with up to $N_s$ UEs, and each UE can only be associated with one SBS. We use a variable $y_{ij,t}$ to indicate user association, which is defined as
\begin{equation}
y_{ij,t} \!=\!
\begin{cases}
1,& \text{if UE $i$ is associated with SBS $j$ at time $t$} \\
0,& \text{otherwise.}
\end{cases}
\label{eq1}
\end{equation}
The number of UEs associated with SBS $j$ at time $t$ is given by $N_{j,t} = \sum_{i \in U} y_{ij,t}$.

In the access downlink, we assume that the antenna arrays of each SBS and its served UE at each time slot perform directivity beamforming before transmission. The
receive power at UE $i$ from SBS $j$ at time ${t}$ can be written as
\begin{equation}
p_{ji,t}^{ac} = {P_{S}}{G_{s}}\left( {j,i} \right){G_{r}}\left( {j,i} \right) \mathcal{L}_t \left( d_{ji} \right), \label{eq3}
\end{equation}
where ${P_{S}}$ is the transmit power at SBS ${j}$, ${G_s}\left( {j,{\rm{\;}}i} \right)$ denotes the transmit antenna gain in the direction of SBS $j$ $\rightarrow$ UE $i$ and ${G_r}\left( {j,{\rm{\;}}i} \right)$ denotes receive antenna gain in the opposite direction, $d_{ji}$ denotes the distance from SBS ${j}$ to UE ${i}$, and $\mathcal{L}_t (d_{ji})$ reflects the path loss and shadow fading of the access link. In this paper, we adopt the close-in free space reference distance path loss model at 28 GHz in~\cite{42b}, which is based on the real propagation measurements at 28 GHz in downtown Manhattan. Therefore, the model can better reflect the propagation characteristics of 28 GHz signals in the real environment. The path loss is expressed as
\begin{equation}
\mathcal{L}_t \left( {{d_{ji}}} \right)\mbox{[dB]} \!=\! 10{\log _{10}}{\!\left( \frac{4\pi {d_0}}{\lambda} \right)^2\!} \!+\! 10{\bar n_{ac}}{\log _{10}}\left( \frac{d_{ji}}{d_0} \right) \!+\! {X_{ac}},
\label{eq4}
\end{equation}
where $\lambda$ denotes the wavelength, $d_0$ is the far field reference distance, ${\bar n_{ac}}$ is the path-loss exponent and ${X_{ac}}$ represents the log-normal shadowing in dB, which is a zero-mean Gaussian random variable with variance ${\sigma_{ac}^2}$. Both ${\bar n_{ac}}$ and ${\sigma_{ac}^2}$ are the best fit over all measurements from the particular measurement campaign at 28 GHz in downtown Manhattan~\cite{42b}.

Unlike sub-6 GHz frequency band, high gain directional link is always established in mmWave communication to compensate for the high path loss and penetration loss of mmWave. Thus, both UEs and SBSs are equipped with antenna arrays for directional beamforming to direct beams towards each other. In this paper, for tractability of the analysis, we utilize a sector antenna model to approximate the actual antenna patterns~\cite{43}:
\begin{align}
	G_{d}(\theta) = \left\{
	\begin{array}{ll}
	G_{d}^{max}, & |\theta| \le \omega_d \\
	G_{d}^{min}, & |\theta| > \omega_d, \\
	\end{array} \right.
	\label{eq4b}
\end{align}
where $d \in \{s,r\}$, $\theta$ denotes the angle off the boresight direction, $\theta \in [-\pi,\pi)$, $\omega_d$ denotes the beamwidth of the main lobe, $G_{d}^{max}$ and $G_{d}^{min}$ is directivity antenna array gain of the main lobe and the side lobes, respectively. We assume that there are no alignment errors in this system, so each intended SBS-UE pair has the maximum directivity gain $G_{SBS}^{max} G_{UE}^{max}$. The beams of other unintended pairs are randomly oriented towards each other and uniformly distributed in $[-\pi,\pi)$. Therefore, the directivity gain of unintended pair is a discrete random variable. The four possible gain values and their corresponding probabilities are given in Table~\ref{antenna}.
\begin{table}[!t]
\centering
\caption{Directivity Gain of Unintended SBS-UE pair $(k,i)$}
\label{antenna}
\begin{tabular}{ll}
\toprule 
${G_{s}}\left( {k,i} \right){G_{r}}\left( {k,i} \right)$ & Probability\\
\midrule 
$G_{s}^{max}G_{r}^{max}$ & $\frac{\omega_s}{2\pi}\frac{\omega_r}{2\pi}$\\
$G_{s}^{max}G_{r}^{min}$ & $\frac{\omega_s}{2\pi}(1-\frac{\omega_r}{2\pi})$\\
$G_{s}^{min}G_{r}^{max}$ & $(1-\frac{\omega_s}{2\pi})\frac{\omega_r}{2\pi}$\\
$G_{s}^{min}G_{r}^{min}$ & $(1-\frac{\omega_s}{2\pi})(1-\frac{\omega_r}{2\pi})$\\
\bottomrule 
\end{tabular}
\vspace{-5mm}
\end{table}

Besides, due to the small wavelength and the directional transmission, the mmWave link is sensitive to the blockage of environmental obstructions like trees, buildings, and human bodies. The most significant difference between the microwave band and the mmWave band is that in many locations, especially when the distance from the transmitter is $>$200 meters, no mmWave signal with transmit powers between 15 and 30 dBm can be detected. It means that all the paths to the receiver in these locations are blocked by obstructions~\cite{45}.

Therefore, we consider the three possible states of the access link, including line-of-sight (LOS), non-line-of-sight (NLOS), and outage~\cite{42b}. The path loss in the outage state is infinite, which means the links between SBS and UE are completely blocked. The access link in the NLOS state will experience more serious path loss than in the LOS state, which is reflected in parameters ${\bar n_{ac}}$ and ${\sigma_{ac}}$ in~(\ref{eq4}). Due to the dynamic and random nature of link blockage, three states of access link are assumed to appear randomly over time, and the path loss of the access link will follow the changes of the access link state at different time $t$~\cite{37}. We use $c_{ji,t}$ to indicate the state of the access link between SBS $j$ and UE $i$ at time $t$, defined as
\begin{align}
	c_{ji,t} \!=\! \left\{
	\begin{array}{ll}
	\!\!0, & \!\mbox{if the state of the access link is LOS} \\
	\!\!0.5, & \!\mbox{if the state of the access link is NLOS} \\
	\!\!1, & \!\mbox{if the state of the access link is outage,} \\
	\end{array} \right.
	\label{eq5}
\end{align}
which can be estimated by SBS $j$ through the statistics of UE $i$ signal. The set of the states of all the possible access links between UE $i$ and all the SBSs can be denoted as ${c_{i,t}} = \left\{ {{c_{1i,t}}, {c_{2i,t}}, ..., {c_{Si,t}}} \right\}$. The probability functions of three states is related to the distance between the transmitting and receiving antennas:
\begin{align}
	& {P_{LOS}}\left( {{d_{ji}}} \right) = (1 - {P_{outage}}\left( {{d_{ji}}} \right)){{\rm{e}}^{ - {a_{los}}{d_{ji}}}} \label{eq7} \\
	& {P_{NLOS}}\left( {{d_{ji}}} \right) = 1 - {P_{outage}}\left( {{d_{ji}}} \right) - {P_{LOS}}\left( {{d_{ji}}} \right) \label{eq8} \\
	& {P_{outage}}\left( {{d_{ji}}} \right) = {\rm{max}}\left( {0,1 - {{\rm{e}}^{ - {a_{out}}{d_{ji}} + {b_{out}}}}} \right). \label{eq9}
\end{align}
where the parameters ${a_{los}}$, ${a_{out}}$, and ${b_{out}}$, and other channel parameters corresponding to the three states can be determined by fitting the equations to the measured data in downtown Manhattan via maximum likelihood estimation~\cite{42b}.

In this system, we assume that TDMA is adopted in each small cell and different frequency bands is allocated for access and backhaul, so no intra-cell interference and backhaul interference exist. However, since small cells are densely deployed and different small cells reuse the access spectrum resources, it is necessary to consider the inter-cell interference. Assume that the bandwidth of the access link is $W_{ac}$, and let $N_0$ denote the one-sided power spectra density of white Gaussian noise. The signal-to-interference-plus-noise ratio (SINR) at UE $i$ associated with SBS $j$ at time ${t}$ can be calculated as
\begin{equation}
\gamma _{ji,t}^{ac} \!=\! \frac{{P_{S}}{G_{s}}\left( {j,i} \right){G_{r}}\left( {j,i} \right) \mathcal{L}_t \left( d_{ji} \right)}{N_0 W_{ac}\!+\! \sum_{k \in B \backslash \{j\} } {P_{S}}{G_{s}}\left( {k,i} \right){G_{r}}\left( {k,i} \right) \mathcal{L}_t \left( d_{ki} \right)},
\label{eq10}
\end{equation}
where interference comes from signals sent from SBSs other than the SBS $j$ associated with UE $i$, and the values of antenna gain ${G_{s}}\left( {k,i} \right){G_{r}}\left( {k,i}\right)$ are shown in Table~\ref{antenna}.

Consider all the UEs associated with the same SBS ${j}$. Then, the average throughput of one of the UEs, UE ${i}$, in the access downlink at time ${t}$
is given by
\begin{equation}
r_{i,t}^{ac} = \sum_{j \in B} \frac{y_{ij,t} W_{ac}}{N_{j,t}} \log_2 \left( 1 + \gamma_{ji,t}^{ac} \right). \label{eq11}
\end{equation}
Since we assume that each UE is associated with one SBS, if UE $i$ is associated with SBS $j^*$, i.e., $y_{ij^*,t} = 1$, the average throughput of UE $i$ in the access link can be written as $r_{i,t}^{ac} =  \frac{W_{ac}}{N_{j^*,t}} \log_2 \left( 1 + \gamma_{j^*i,t}^{ac} \right)$.

\subsubsection{Backhaul Downlink Model}

The backhaul transmissions are performed in 73 GHz. We assume that the MBS can establish directional backhaul connections with all SBSs at the same time. Each SBS is assumed to be allocated orthogonal frequency resources without interference between each other.

In the backhaul downlink, the antenna arrays of each MBS and all the SBSs perform directivity beamforming before transmission. The received power at SBS $j$ at time $t$ is
\begin{equation}
p_{Mj,t}^{bk} = {P_M}{G_s}\left( {M,j} \right){G_r}\left( {M,j} \right){\mathcal{L}_t}\left( {{d_{Mj}}} \right), \label{eq12}
\end{equation}
where ${P_M}$ is the transmit power of the MBS; ${G_s}\left( {M,j} \right)$ and ${G_r}\left( {M,j} \right)$ denote the transmit and receive directivity antenna gain, respectively, from the MBS to SBS $j$, $d_{Mj}$ is the distance from the MBS to SBS $j$; and $\mathcal{L}_t \left( d_{Mj} \right)$ reflects the path loss and shadow fading of the backhaul link. We adopt the close-in free space reference distance path loss model at 73 GHz in~\cite{42b}, which is based on the real propagation measurements at 73 GHz in downtown Manhattan. The path loss is given by
\begin{equation}
\mathcal{L}_t \left( {{d_{Mj}}} \right)\mbox{\![dB]} \!=\! 10{\log _{10}}{\left( \frac{4\pi {d_0}}{\lambda} \right)^2} \!+ 10{\bar n_{bk}}{\log _{10}}\left( \frac{d_{Mj}}{d_0} \right) + {X_{bk}},
\end{equation}
where ${\bar n_{bk}}$ is the path-loss exponent of the backhaul link, and ${X_{bk} \sim N(0, \sigma^2_{bk})}$ is related to shadow fading in the backhaul link. Both ${\bar n_{bk}}$ and ${\sigma_{bk}^2}$ are the best fit over all measurements from the particular measurement campaign at 73 GHz in downtown Manhattan~\cite{42b}. Considering the high gain directional backhaul link achieved by beamforming at antenna arrays, we adopt the same sector antenna in the backhaul link as in the access link. In addition, as mentioned before, we assume that with the appropriate deployment of BSs, the backhaul links can be regarded as LOS connections.

The total bandwidth of the backhaul network is $W_{bk}$. We assume that MBS allocates orthogonal frequency resources to each SBS, and we denote the bandwidth allocated by the MBS to the SBS $j$ at time $t$ as $w_{j,t}$. In this paper, in order to allocate the backhaul resources more effectively, we consider the available backhaul resources for each UE. In this way, the allocation of backhaul resources can better adapt to the dynamic changes of access link state. Accordingly, the backhaul resources of each SBS, i.e., the bandwidth of each backhaul link, is the sum of the backhaul resources of all the associated UEs. If UE $i$ is associated with SBS $j$, the proportion of the backhaul bandwidth that is allocated to UE $i$ at time ${t}$ is given by the backhaul bandwidth factor $\beta_{ij,t} \in \left[ 0, 1 \right]$, and the amount of bandwidth assigned to UE $i$ is $\beta_{ij,t} W_{bk}$. So the total proportion of the backhaul bandwidth allocated to SBS $j$ at time $t$ is $\beta_{j,t} = \sum_{i \in U} y_{ij,t} \beta_{ij,t}$, and the total amount of bandwidth for SBS $j$ is $w_{j,t} = \beta_{j,t} W_{bk}$.

In the backhaul link, due to the different frequency bands used for access and backhaul and the orthogonal resource allocation, we don't need to consider the access interference and the interference between SBSs. Besides, we consider the single MBS in our system, and assume the sufficient long distances between the SBSs in the HetNet with the neighbouring MBSs. So the signal-to-noise ratio (SNR) at SBS $j$ at time $t$ can be calculated as
\begin{equation}
\gamma_{j,t}^{bk} = \frac{{P_M}{G_s}\left( {M,j} \right){G_r}\left( {M,j} \right){L_t}\left( {{d_{Mj}}} \right)}{N_0 w_{j,t}}. \label{eq15}
\end{equation}
Then, the throughput of backhaul link between MBS and SBS $j$ at time ${t}$ is given by $r_{j,t}^{bk} = w_{j,t} {\log _2}\left( {1 + \gamma _{j,t}^{bk}} \right)$. And the achieveable throughput of UE $i$ in the backhaul link at time $t$ is given by
\begin{equation}
r_{i,t}^{bk} = \mathop \sum \limits_{j \in B} {y_{ij,t}}{\beta _{ij,t}}{W_{bk}}{\log _2}\left( {1 + \gamma _{j,t}^{bk}} \right). \label{eq17}
\end{equation}
Since we assume that each UE is associated with one SBS, if UE $i$ is associated with SBS $j^*$, i.e., $y_{ij^*,t} = 1$, the average throughput of UE $i$ in the backhaul link can be written as $r_{i,t}^{bk} =  {\beta _{ij^*,t}}{W_{bk}}{\log _2}\left( {1 + \gamma _{j^*,t}^{bk}} \right)$.

\vspace{-2mm}
\subsection{Problem Formulation}\label{S3-2}

We consider both the access link and backhaul link. At time $t$, in order to serve UE $i$, the actual achievable link throughput in the HetNet is determined by the small one between the backhaul link throughput and the access link throughput, which is given by ${R_{i,t}} = {\rm{\;min}}\left( {r_{i,t}^{bk},r_{i,t}^{ac}} \right)$. The total actual link throughput of all the UEs at time $t$ in the HetNet is $R_t = \sum_{i \in U} R_{i,t}$.

In the mmWave HetNet, we aim to maximize the long-term total throughput of all the UEs over a finite period $T$. The joint user association and backhaul resource allocation problem can be formulated as
\begin{align}
&\max_{y, \beta} \,\, \sum\limits_{t=1}^T{R_t}  \label{eq20} \\
& s.t. \quad\sum_{j \in B} y_{ij,t} = 1, \; \forall i \in U, \; \forall t \label{eq21} \\
&\; \qquad \, \sum_{j \in B} \beta_{j,t} \le 1, \; \forall t \label{eq22} \\
&\; \qquad \, N_{j,t} = \sum_{i \in U} y_{ij,t} \le {N_s}, \; \forall j \in B, \; \forall t  \label{eq23} \\
&\; \qquad \, y_{ij,t} \in \{ {0,1} \}, \; \forall j \in B, \; \forall i \in U \label{eq24} \\
&\; \qquad \, \beta_{ij,t} \in \left[ {0,1} \right], \; \forall j \in B, \; \forall i \in U. \label{eq25}
\end{align}

This joint optimization problem can be solved by finding the optimal backhaul bandwidth factor set $\beta$ and the user association indicator set $y$. Due to the dynamically changing blockage state of the access links, the user association and backhaul resource allocation strategy need to be adjusted in time in order to ensure good throughput performance. Constraint~(\ref{eq21}) indicates that each UE can only be associated with one SBS at a time. Constraint~(\ref{eq22}) indicates that the bandwidth resources shared by all backhaul links does not exceed the total available bandwidth resources. Constraint~(\ref{eq23}) ensures that the number of UEs served by each SBS does not exceed the upper limit of the SBS.

Problem~\eqref{eq20} is a MINLP problem, which is difficult to solve for the global optimal solution. Many state-of-the-art approaches are helpful to solve the MINLP problem, including heuristic-based, optimization-based, and game theory-based approaches~\cite{26}. However, in our system, the access link state changes over time, such approaches will not be accurate. They need to reconfigure to reflect the new environment. Besides, most of these approaches rely on perfect knowledge of wireless environment or a large amount of information exchanged between network entities (e.g., BSs and UEs). They are difficult to perform in the large-scale HetNets.
Therefore, we propose a MADRL-based method to solve these problems. The method allows each agent to learn to adapt to the environment without the knowledge of environment in advance and locally find the optimal policy through effective learning. Table~\ref{Notation} summarizes the notations adopted in this section.

\begin{table}[!t]
\caption{Notation Summary}
\vspace{-3mm}
\label{Notation}
\centering
\begin{tabular}{ll}
\toprule 
Notation & Description\\
\midrule 
$S$, $N$ & number of SBSs, UEs\\
$B$, $U$  & the set of SBSs, UEs\\
$y_{ij,t}$, $N_{j,t}$  & user association indicator\\
$\beta_{ij,t}$, $\beta_{j,t}$, $w_{j,t}$  &  backhaul bandwidth allocation indicator\\
$p_{Mj,t}^{bk}$, $p_{ji,t}^{ac}$ & receive power \\
$P_S$, $P_M$ & transmit power\\
$d_{ji}$, $d_{Mj}$ & distance between transceivers\\
$\mathcal{L}_t \left( \cdot \right)$ & large scale channel gain\\
${\bar n_{ac}}$, ${\bar n_{bk}}$ & path-loss exponent\\
${X_{ac}}$, ${X_{bk}}$ & log-normal shadowing\\
$\sigma_{ac}$, $\sigma_{bk}$ & variance of the shadow fading\\
$G_s(\ast,\cdot)$ & transmit antenna gain from $\ast$ to $\cdot$\\
$G_s^{max}$, $G_s^{min}$ & transmit antenna gain value\\
$\omega_s$, $\omega_r$ & main lobe beamwidth\\
$G_r(\ast,\cdot)$ & receive antenna gain from $\ast$ to $\cdot$\\
$G_r^{max}$, $G_r^{min}$ & receive antenna gain value\\
$c_{ji,t}$, $c_{i,t}$ & access link state indicator\\
$W_{ac}$, $W_{bk}$ & bandwidth\\
$\gamma _{ji,t}^{ac}$, $\gamma_{j,t}^{bk}$ & SINR for access, SNR for backhaul\\
$r_{i,t}^{ac}$, $r_{j,t}^{bk}$, $r_{i,t}^{bk}$ & link throughput\\
$R_{i,t}$, $R_{t}$ &  actual achievable link throughput\\
\bottomrule 
\end{tabular}
\vspace{-8mm}
\end{table}

\section{Multi-agent Deep Reinforcement Learning based Method}\label{S4}
In this paper, we propose a MADRL based method to solve problem~\eqref{eq20}. Note that it is not practical to use single-agent deep reinforcement learning (SADRL) method for this problem. On the one hand, the state and action space will grow dramatically, which makes the convergence of the algorithm extremely difficult. Specifically, if we denote the size of action space for each agent of the MADRL method as $\mathcal{A}$, the size of action space for the agent of the SADRL method is $\mathcal{A}^{N}$. Thus, we can see that the growth is even more significant in the HetNets we considered with small cells densely deployed. On the other hand, the SADRL method needs a network controller with agent to collect information from all the network entities, make decisions centrally and then broadcast the decisions to each network entity. The considerable communication overhead will be intolerable in large-scale HetNets. In comparison, the MADRL method we proposed supports that each agent obtain state observations independently and make decision locally, which greatly reduces communication overhead. Besides, the MADRL method we proposed trains all the agents in parallel, thus reducing the training time.

In this section, we first establish a Markov game model corresponding to the joint optimization problem. Then we introduce DDQN algorithm and develop a multi-agent double deep Q-learning (MADDQN) method to obtain the optimal solution for the Markov game model.
\vspace{-3mm}
\subsection{Markov Game Model}\label{S4-1}
\begin{figure}[!t]
\centering
\includegraphics*[width=3.0in]{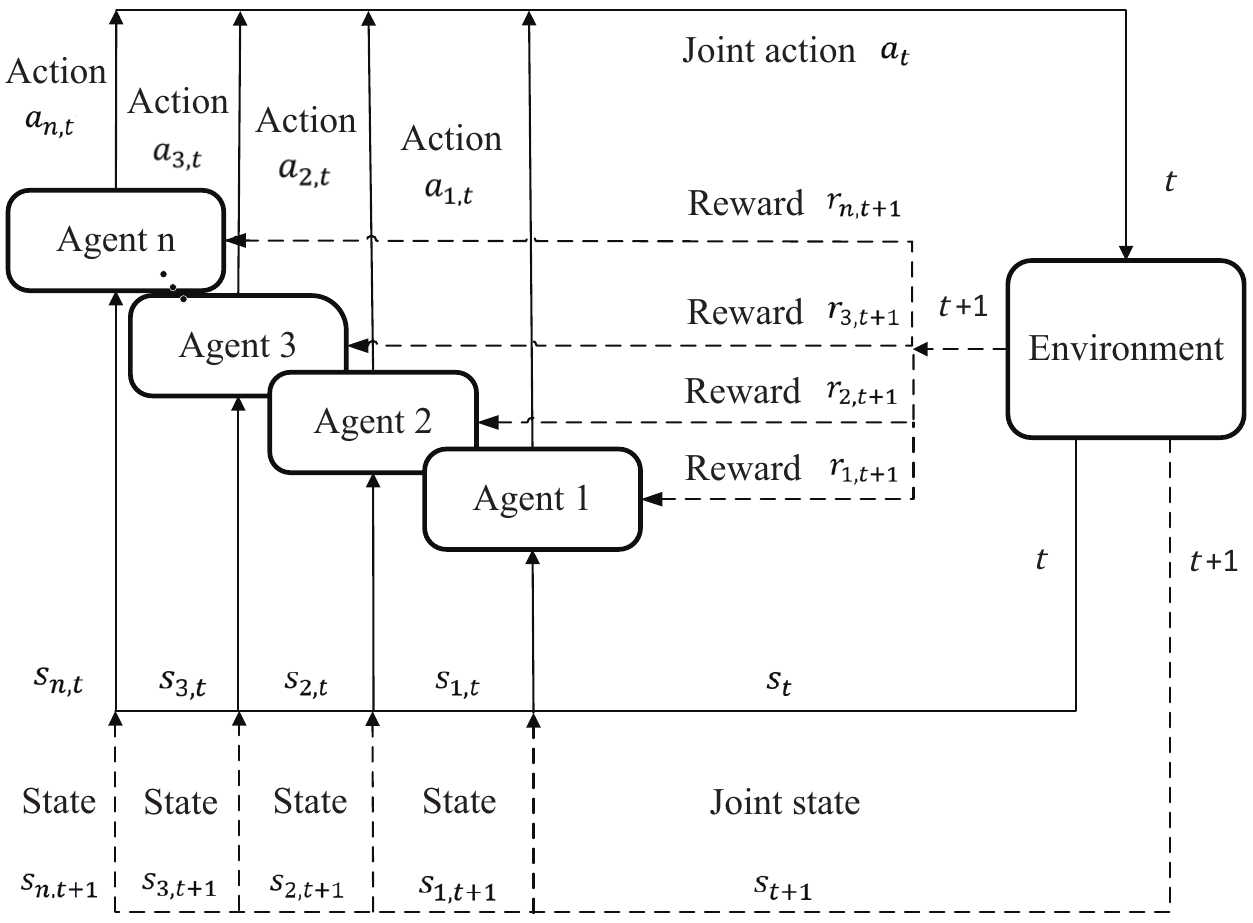}
\centering
\caption{The Markov game model.}
\label{fig:2}
\vspace{-5mm}
\end{figure}

In MADRL, the interaction between agents and the environment is usually described as a Markov game~\cite{23}, as shown in Fig.~\ref{fig:2}. There are four key elements in a Markov game: (i) a finite environment state space $S$, (ii) a finite action space $A$, (iii) the reward $R$, and (iv) the state transition probability $P$. Suppose that there are $n$ agents in the environment. At training time $t$, each agent $i$ observes the current environment state $s_{i,t}$, and then takes an action $a_{i,t}$. The joint actions of all the agents in the environment are represented by $a_{t}$. Afterwards, the environment feeds back the reward for each agent according to the joint actions, while the state transition occurs simultaneously. Each agent $i$ then receives a reward $r_{i,t+1}$ and observes a new environmental state $s_{i,t+1}$. Then the interaction between each agent and the environment at training time $t$ is completed. The state transition probability and reward at time $t+1$ depend only on the previous state and the previous action at time $t$, regardless of the earlier states and actions (i.e., satisfying the Markov condition).

Since machine learning algorithms are computationally and memory intensive, some notable efforts have already been made both in hardware design and software acceleration, which makes it possible to move the optimization process at the UE~\cite{45b,45c,45d}. In this sense, \cite{39} and \cite{41} have treated UEs as agents and proposed distributed MARL algorithms to solve the user association problem. Thus, in this system, we regard each UE $i$ as an agent.

The entire mmWave HetNet system can be regarded as the environment. At each time $t$, after the user association and backhaul resource allocation strategies of all the UEs are adopted, the link throughput corresponding to each UE and the total link throughput can be obtained from the environment. This can be used as the basis for UEs' decision-making.

To transform the joint optimization problem into a Markov game, which can be solved by MADRL, we design the key elements of the
corresponding Markov game below in detail.

\smallskip
\subsubsection{\textbf{State}}
State should represent the features of the network environment at different times, when different resource allocation and user association policies lead to different link throughput. Moreover, due to the dynamic characteristics of mmWave links, the access links assume different states, which change over time. Therefore, the state should contain the total achieved link throughput in the HetNet, the achieved link throughput of each UE and the states of all the access links.

Besides, considering the constraint~(\ref{eq22}), since each UE makes decisions independently, the total backhaul resources allocated to UE may exceed the limit. Thus, each UE needs to observe the allocation of backhaul resources in the network, which is determined by actions of all the UEs at previous time. We can use $\beta_{t}$ to represent the observation of the backhaul resource allocation at time $t$, which is given by
\begin{equation}
\beta_{t} =
\begin{cases}
\sum_{j \in B} \beta_{j,t}, & \text{$\sum_{j \in B} \beta_{j,t} \le 1$} \\
0, & \text{otherwise.}
\end{cases}
\label{eq26}
\end{equation}

Therefore, for each UE $i$, the state observation obtained from the environment consists of four parts:
(i) the total achieved link throughput, (ii) the observation of the backhaul resource allocation, (iii) the achieved link throughput of UE $i$ and (iv) the states of all the possible access links of UE $i$.

In addition, in MADRL,  since all the agents learn to select their actions at the same time, each agent faces a non-stationary environment, which is harmful to the experience replay in DQN. Therefore, we adopt the fingerprint-based method designed in~\cite{46} to deal with this problem. The main idea is to add the estimate of other agents' policies to the state space of each agent. However, the policy of each agent includes a high-dimensional DQN, which makes it difficult to act as a part of state. As for MADRL, the policy updates of each agent are correlated with the number of training iterations, denoted by $e$, and the exploration rate $\varepsilon$ in the $\varepsilon$-greedy strategy. Therefore, these two variables can be added to the state space of each agent as low dimensional fingerprints to track the historical trajectory of other agents' policy.

As a result, the state observation of each agent $i$ at time $t$ can be designed as ${s_{i,t}}\!\! =\!\! \left\{ \beta_{t-1}, R_{t}, c_{i,t}, R_{i,t}, e, \varepsilon \right\}$, and the joint actions of all the agents can be expressed as ${s_t}\! =\! \left\{ s_{1,t}, s_{2,t}, ..., s_{N,t} \right\}$.

\smallskip
\subsubsection{\textbf{Action}}
The action of each UE $i$ consists of two parts: (i) backhaul bandwidth allocation, and (ii) user association. However, the backhaul bandwidth factor indicating the bandwidth allocation, i.e., $\beta_{ij,t}$,
is a
fraction
in [0,1],
leading to
a continuous action space, which the DQN algorithm is not good at dealing with. Therefore,
discretization of
the
action space
should be
considered.

We divide the total backhaul bandwidth into $L$
non-overlapping
bandwidth resource blocks.
Thus
the backhaul bandwidth allocation problem is transformed into a bandwidth resource block allocation problem. Suppose that
each
bandwidth resource block can only be allocated to one UE,
while
a UE can occupy multiple bandwidth resource blocks. At time $t$, if UE $i$ is associated with SBS $j$, the number of backhaul bandwidth resource blocks that can be occupied by UE $i$ is represented
by
$l_{ij,t}$. Therefore, the backhaul bandwidth factors $\beta_{ij,t}$ can be written as $\beta_{ij,t} = \frac{l_{ij,t}}{L}$.
Accordingly, the constraint~(\ref{eq22}) is equivalent to $\sum_{j \in B}\sum_{i \in U} l_{ij,t} \le L, \; \forall t$.

Then, we can use $l_{ij,t}$ to represent the allocation of backhaul bandwidth
to
UE $i$ at time $t$. In order to reduce the action space and avoid
the case when
the
backhaul bandwidth
is allocated to only one or
few
UEs,
we set an
upper limit
on the available
bandwidth resource blocks for each UE
according to the number of UEs in
the HetNet,
denoted by $l_{max}$.

As a result, the action of each agent $i$ at time $t$ is designed as $a_{i,t} = \left\{ j^*, l_{ij^*,t} \right\}$, where ${j^*}$ denotes the index of the SBS associated with UE $i$, and $l_{ij^*,t}$ satisfies $l_{ij^*,t} \in \left[ 0, l_{max} \right]$. Therefore, the size of the action space, denoted by $\mathcal{A}$, is the product of the size of the range $\left[ 0, l_{max} \right]$ and the number of SBSs in the HetNet, i.e., $(l_{max} + 1) \times S$. The joint actions of all the agents are expressed as $a_t = \left\{ a_{1,t}, a_{2,t}, ..., a_{N,t} \right\}$.

\smallskip

\subsubsection{\textbf{Reward}}
In the HetNet, all the agents make bandwidth allocation and user association decisions in order to maximize the long-term total link throughput. However, in the training process, the total allocated backhaul resources may exceed the upper limit. At this time, constraint~(\ref{eq22}) is not satisfied. In addition to add the observation of bandwidth allocation to the state mentioned before, it is useful to constrain the reward value. If the total allocated backhaul resources are more than the upper limit, the reward value is equal to 0. This means that, if the joint actions of all the agents do not meet the constraint~(\ref{eq22}), all the agents will not be rewarded. Besides, if the number of UEs served by SBS $j$ is more than $N_s$, which means the constraint~(\ref{eq23}) is not satisfied, the UEs associated with SBS $j$ will not be rewarded. Therefore, agents can learn to avoid constraint violations in the training process.

Moreover, it seems like all the agents assume a common goal that maximize the long-term total link throughput. However, this cannot be simply regarded as a fully cooperative MARL~\cite{47}. Each UE has the selfishness to improve its own throughput. Therefore, in the design of reward value, we need to consider not only the total link throughput in the network, but also the link throughput of each UE. We define $\delta$ as a factor representing the degree of selfishness of each UE. The reward of each agent $i$ is written as
\begin{equation}
r_{i,t+1} \!=\!
\begin{cases}
\delta R_{i,t}\! +\! (1\!-\!\delta) \frac{R_t}{N}, & \text{\!\!\!if $\beta_{t} \!\le \!1$ and $N_{j^*\!,t}\! \le \!N_s$ } \\
0, & \text{\!\!\!otherwise.}
\end{cases}
\label{eq32}
\end{equation}
The larger the value of $\delta$, the more likely UEs are to make decisions conducive to improving their own throughput.
We will determine $\delta$ with the best network performance through simulation experiments.

\vspace{-3mm}
\subsection{Double Deep Q-Learning (DDQN) Algorithm}\label{S4-2}

To solve the Markov game, it is necessary to select appropriate policy-making algorithm for each agent. The policy $\pi$ refers to the mapping from state $s$ to action $a$, which defines the behavior of agent. The policy is expressed by the probability of taking action $a$ in the current state $s$. Considering long-term reward, the goal of the agent is to find an optimal policy that maximizes the cumulative discounted reward $G_t$ at each training step $t$, which is given by $G_t \doteq \sum_{k = 0}^\infty \gamma^k r_{t + k + 1}$, where $\gamma$ is the discount rate, which reflects the importance of future rewards, and $r_t$ is the reward value at training step $t$.

To evaluate the policies, the action-value function, a.k.a. the Q-value, is adopted in Q-learning. It is defined as the expected return of the discounted reward of all possible policy sequences after taking action $a$ in the current state $s$ at training step $t$ according to policy $\pi$, which can be written as
\begin{equation}
Q(S_t, A_t) = \mathbb{E}_\pi \left[ G_t \;|\; S_t = s, A_t = a \right]. \label{eq34}
\end{equation}

Following the Bellman equation~\cite{20}, the current action-value function at training step $t$ can be associated with the subsequent action-value function at training step $t+1$. The optimal strategy corresponds to the optimal action-value function in the finite Markov decision process (MDP)~\cite{20}. Consequently, the optimal action-value function, which is also called the target Q-value, can be written as $Q_{target} (S_t, A_t) = r_{t + 1} + \gamma \cdot \max _a Q( S_{t + 1}, a)$. The Q-value is updated with the target Q-value at each training step $t$, denoted as
\begin{equation}
Q(S_t, A_t) := Q(S_t, A_t) + \alpha \cdot [ Q_{target}(S_t, A_t) - Q(S_t, A_t) ], \label{eq36}
\end{equation}
where $\alpha$ represents the learning rate, which determines the updates of the Q-value.

Q-learning needs to establish a Q-table to store the Q-values of all the state-action pairs. For high-dimensional state and action spaces, the Q-table will be very large, which brings a great burden on storage and computing. To address this issue, DQN leverages a DNN to learn and approximate Q-values, which is referred to as the Q-evaluate network. Nevertheless, DNN needs a large number of labeled data for training, while RL has to generate the training data in the learning process, i.e., it does not provide a labeled dataset in advance. In addition, the training data should be uncorrelated, but RL obtains highly correlated data during its operation, which could cause the training process to be unstable. Therefore, \emph{experience replay} and \emph{fixed Q-target network} are used to improve the stability of the training process~\cite{22}.

\emph{Experience replay }refers to storing the experience obtained from interaction with the environment at each training step $t$, i.e.,  $\{S_{t}, A_{t}, r_{t+1}, S_{t+1}\}$, in the replay memory. Then, when the Q-evaluate network needs to be updated, a mini-batch of replay memory $D$ is randomly selected from the replay memory as training data, both the new data and historical data will be included for training, thus breaking the correlation in the data.

\emph{Fixed Q-target network} means the use of an independent Q-target network to generate the target Q-value. The update of the Q-target network is slower than that of the Q-evaluate network, and is achieved by replacing the parameters of the Q-target network $\theta^-$ with the parameter of the Q-evaluate network $\theta$.

Therefore, unlike the update of Q-table in Q-learning, as in~(\ref{eq36}), DQN updates the parameter $\theta$ of the Q-evaluate network using mini-batches randomly selected from the replay memory to minimize the following loss function.
\begin{equation}
L = \sum_D \left[ r_{t+1} + \gamma \cdot \max_a Q(S_{t+1}, a; \theta^-) - Q(S_t, A_t; \theta) \right]^2. \label{eq37}
\end{equation}

Moreover, when calculating the target Q-value, both the selection and the evaluation of actions based on the maximum Q-value are estimated by the Q-target network, which is prone to overestimation. To this end, DDQN decouples the selection and evaluation to solve the above problem~\cite{48}. In the calculation of the target Q-value, action selection uses the Q-value estimated by the Q-evaluate network, and the Q-target network is only used to evaluate actions. The loss function of the DDQN is modified as
\begin{align}
L =&\; \sum_D [ r_{t+1} + \gamma \cdot Q(S_{t+1}, \argmax_a Q(S_{t+1}, a; \theta); \theta^-)
\nonumber \\
&\; - Q(S_t, A_t; \theta) ]^2.
\label{eq38}
\end{align}

\vspace{-4mm}
\subsection{Multi-Agent Double Deep Q-Learning (MADDQN) Method }\label{S4-3}

Taking DDQN as the policy-making algorithm of each agent, we design MADDQN method to solve the Markov game model. We assume the Markov game model is episodic. Each episode includes $T$ steps. Each step $t$ corresponds to the time $t$. The state of the access link between each UE and each SBS changes over training time steps, so that agents can learn the link blockage pattern and adaptly make optimal decisions. Our goal is to maximize the sum of the total link throughput in an episode, corresponds to~(\ref{eq20}).

In this paper, we mainly consider the design of joint optimization scheme that can effectively cope with random and dynamic link blockage. In fact, the network may change in other aspects at the same time, such as the rapid movement of UEs and the increase or decrease of the number of UEs, etc. If we spend too much time for training, these changes in the network during the training stage may affect the results of the training. Thus, we adopt a distributed architecture to design algorithm training, each UE can train its Q network locally. From the perspective of the entire network, training is done in parallel, greatly reducing the training time. Besides, the increase in computing power is rapid, so the time cost of training models will be greatly reduced in the future.

The detailed training procedure is presented in Algorithm~\ref{alg.a}. The agent at each UE $i$ has two dedicated DQNs: the Q-evaluate network and the Q-target network. At each episode, all the agents' states are first initialized. The UE-SBS association has not been established in the initial state, so the link throughput of all the agents are zero at this time. Then, at each training step $t$, after each UE agent observes the environment state $s_{i,t}$, the Q-evaluate network uses it as input and outputs the Q-values of all the state-action pairs in the current state. The choice of action is based on the Q-values with the $\varepsilon$-greedy policy, which means that action is randomly chosen with probability $\varepsilon$ while the action with the maximum Q-value is chosen with probability $(1 - \varepsilon)$.

\begin{algorithm}[htbp]
\caption{The MADDQN Method for Joint Optimization of User Association and Resource Allocation in the HetNet}	
\label{alg.a}
\begin{algorithmic}[1]
\STATE Initialize the parameters of the Q-evaluate network and the Q-target network of all the UE agents randomly;
\STATE Initialize the replay memory at each agent;
\FOR{each episode $e$}
\STATE Initialize $s_t$;
\FOR{each step $t$}
\FOR{each agent $i$}
\STATE Observe environment state $s_{i,t}$;
\STATE The Q-evaluate network uses $s_{i,t}$ to choose action $a_{i,t}$ from $\mathcal{A}$ with the centralized $\varepsilon$-greedy policy;
\ENDFOR
\STATE All agents take actions and obtain reward $r_{i,t+1}$;
\STATE Update access link states $c_{ji,t}$;
\FOR{each agent $i$}
\STATE Observe environment state $s_{i,t+1}$;
\STATE Store $\{s_{i, t}, a_{i, t}, r_{i, t+1}, s_{i, t+1}\}$ in replay memory;
\ENDFOR
\ENDFOR
\FOR{each agent $i$}
\STATE Sample minibatch ${D_i}$ from replay memory randomly;
\STATE Calculate the loss function using Q-target network;
\STATE Update the parameters of the Q-evaluate network using stochastic gradient descent;
\STATE For every $C$ episodes, update the parameters of the Q-target network;
\ENDFOR
\ENDFOR
\end{algorithmic}
\end{algorithm}

However, if all the agents randomly select $l_{ij,t}$ between $\left[ 0, l_{max} \right]$, the number of total backhaul resource blocks allocated to UEs in the network may be all the values between 0 and $Nl_{max}$. The probability of $L$ resource blocks being fully utilized is only $\frac{1}{Nl_{max}+1}$. Actually, we hope that $L$ resource blocks can be utilized as much as possible, and $L$ should be the maximum number of resource blocks available. Therefore, we implement $\varepsilon$-greedy strategy centrally. Specifically, all the UEs are consistent in the mode of selecting actions, i.e., randomly or according to the output of Q-evaluate network. The mode is controlled by a network controller located in the MBS.

If the network controller decides to select actions according to the output of the Q-evaluate network, all the UEs can make action selection decisions independently according to their Q-evaluate network. Then, each UE sends the access request and the number of required backhaul resources to the SBS that the UE has chosen. If the number of UEs associated with the SBS satisfies constraint~(\ref{eq23}), the SBS accepts the access request and sends a feedback signal to the UE to establish an access connection. Otherwise, the SBS rejects the access request and the access throughput of all associated UEs is 0. Then, all the SBSs send the total backhaul resource requirements of the associated UEs to the MBS. The MBS determines whether the total backhaul resource allocation satisfies the constraint~(\ref{eq22}). If satisfied, the MBS allocates the corresponding backhaul resources to SBSs and feeds back the information of the allocated total backhaul resources to each UE. If not, the backhaul resources will not be allocated. At this time, the throughput of all the UEs is 0.

If actions are decided to be selected randomly, the network controller randomly generates rational action selection for each UEs. Specifically, the user association policy should meet the constraint~(\ref{eq21}), while the backhaul resource allocation policy should meet $\sum_{j \in B}\sum_{i \in U} l_{ij,t} = L$, so as to make sure all the backhaul resources can be fully utilized. The MBS can directly determine the allocation of the backhaul resource based on the action selection and send the action selection scheme to each SBS. Then, each SBS sends the action selection scheme to the previous associated UEs. After receiving the new scheme, each UE sends an access request to the corresponding SBS. The SBS directly accepts the request and sends a feedback signal to the UE to establish an access connection.

After all the agents take actions at each training step $t$, the achievable link throughput of each UE can be observed locally. Each UE can send its achievable link throughput to the MBS through the associated SBS. The MBS will calculate the current total achievable link throughput, and then feed it back to each UE, which makes each UE can calculate its complete reward $r_{i,t + 1}$. When the access link state of UE $i$ changes, each SBS $j$ estimates its current access link state $c_{ji,t}$, and sends the state to UE $i$. It should be noted that taking into account the blockage characteristics of the mmWave link, we use the sub-6 GHz band to reliably transmit access requests and signaling information. Therefore, we can see that through information exchange, the state observations of each UE at step $t+1$ can be completely obtained, which can be used as the basis for action selection at step $t+1$. Then, the tuple $\{s_{i,t}, a_{i,t}, r_{i,t + 1}, s_{i,t + 1}\}$ of each UE $i$ is stored in its replay memory for the update of the Q network.

\begin{table}[!t]
\caption{Simulation Parameters}
\vspace{-3mm}
\label{Simulation}
\centering
\begin{tabular}{ll}
\toprule 
Parameter & Value\\
\midrule 
MBS transmit power ${P_0}$ & 40 dBm\\
SBS transmit power ${P_j}$ & 30 dBm\\
Backhaul total bandwidth ${W_{bk}}$ & 6 GHz\\
Number of resource blocks  ${L}$ & 300\\Access total bandwidth ${W_{ac}}$ & 1 GHz\\
Noise power density ${N_0}$ & -114 dBm/MHz\\
Maximum number of UEs served by a SBS ${N_s}$ & 20\\
Main lobe gain $G_{MBS}^{max}$, $G_{SBS}^{max}$, $G_{UE}^{max}$ & 20dB, 10dB, 5dB\\
Side lobe gain $G_{MBS}^{min}$, $G_{SBS}^{min}$, $G_{UE}^{min}$ & -10dB, 0dB, 0dB\\
Main lobe beamwidths $\omega_{MBS}$, $\omega_{SBS}$, $\omega_{UE}$ & $30^{\circ}$, $40^{\circ}$, $50^{\circ}$\\
Access channel parameters ${\bar n_{ac}}$, ${\sigma _{ac}}$ (LOS)  & 2.1, 3.6 dB\\
Access channel parameters ${\bar n_{ac}}$, ${\sigma _{ac}}$ (NLOS)  & 3.4, 9.7 dB\\
Backhaul channel parameters ${\bar n_{bk}}$ and ${\sigma _{bk}}$ & 2.0, 4.2 dB\\
Link state parameters $\frac{1}{a_{out}}, {b_{out}}$, and $\frac{1}{a_{los}}$ & 50 m, 1.8, 50 m\\
\bottomrule 
\end{tabular}
\vspace{-5mm}
\end{table}

Furthermore, at the end of each episode $e$, a mini-batch of memory $D_i$ is randomly sampled from the replay memory as training data for the Q-evaluate network at each UE agent $i$. With the target Q-value calculated by the Q-target network, the loss function of the Q-evaluate network can be written as
\begin{align}
L_i =&\; \sum_{D_i} [ r_{i, t+1} + \gamma \cdot Q(s_{i,t+1}, \argmax_a Q(s_{i, t+1}, a; \theta_t); \theta_t^-)
\nonumber \\
&\; - Q(s_{i,t}, a_{i,t}; \theta_t) ]^2.
\label{eq40}
\end{align}
Stochastic gradient descent is used to update the Q-evaluate network. The Q-target network of each UE agent $i$ is updated in every $C$ episodes according to the fixed Q-target.

After completing training of the multi-agent system, the trained Q-evaluate network of each UE $i$ can be used to make user association and backhaul bandwidth allocation decisions in the current system scenario. At this time, each agent $i$ can independently choose action without the assistance of the network controller. Specifically, when the access link state changes, each UE agent $i$ uses the state observation with $e$ and $\varepsilon$ from the last training episode as input to the Q-evaluate network. According to the output of the Q-evaluate network, the action with the maximal Q-value is chosen by each UE. The information exchange in this process is similar to the Q-value based action selection in the training stage. Thereafter, at each time $t$, all the UEs will be associated with the corresponding SBS and the corresponding backhaul bandwidth will be allocated to each SBS.

\vspace{-2mm}
\section{Simulation Results and Discussions}\label{S5}

In this section, we evaluate the performance of the proposed MADDQN method through simulations. Specifically, we first investigate the impact of the selfish factor $\delta$ on the performance of our scheme. Then, we compare our scheme with the other four schemes in terms of the throughput performance.

\begin{figure}[t]
\centering
\includegraphics*[width=2.4in]{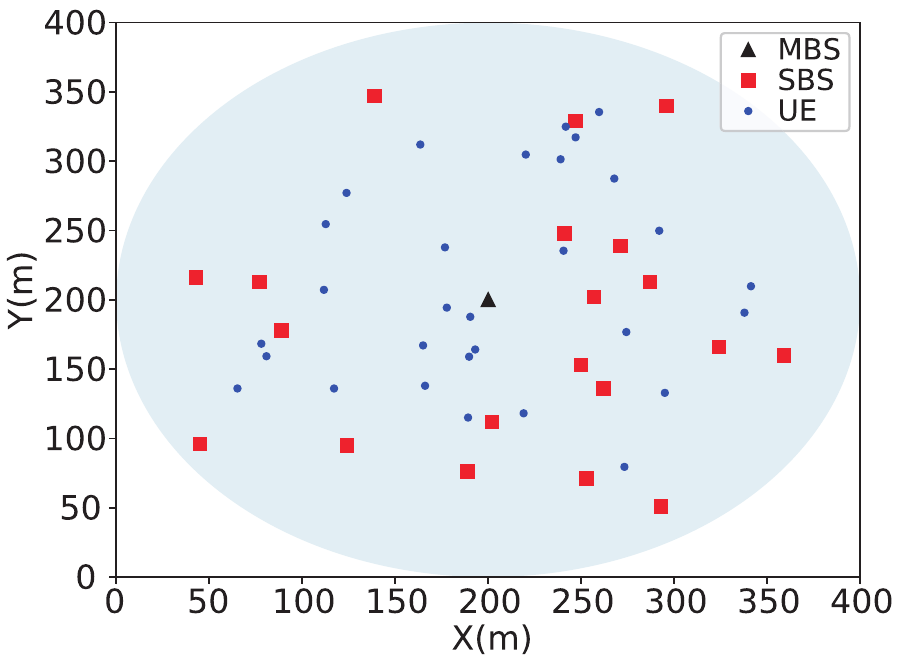}
\vspace{-3mm}
\caption{The deployment scenario of the HetNet used in our simulation study.}
\label{fig:3}
\vspace{-5mm}
\end{figure}

\vspace{-2mm}
\subsection{Simulation Setup }\label{S5-1}
\vspace{-1mm}

\begin{figure*}[t]
\label{fig:mini:subfig:a}
\begin{minipage}[t]{0.19\linewidth}
\centering
\includegraphics[width=1\columnwidth,height=1.1in]{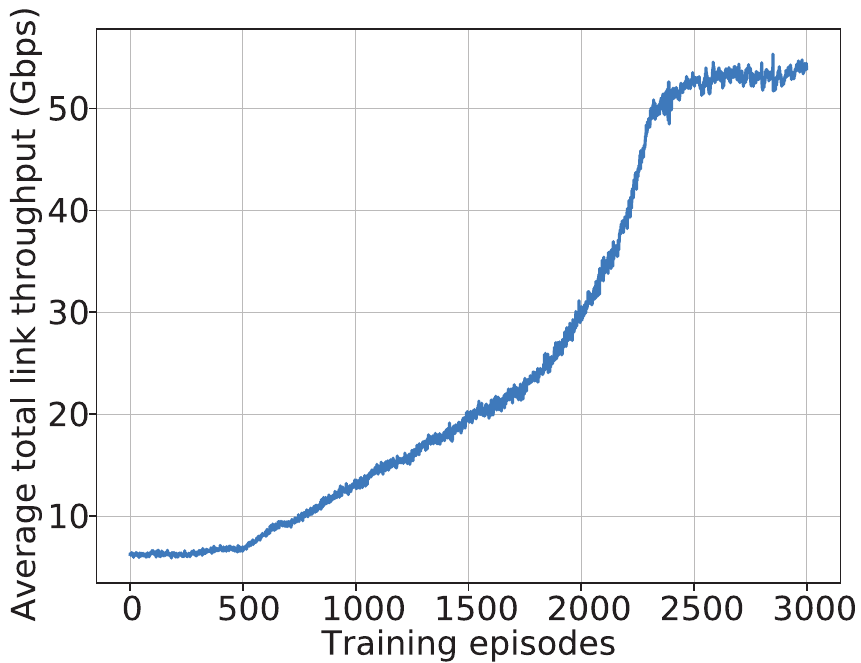}
\centerline{\quad\small (a) $\delta=0$}
\end{minipage}%
\label{fig:mini:subfig:b}
\begin{minipage}[t]{0.19\linewidth}
\centering
\includegraphics[width=1\columnwidth,height=1.1in]{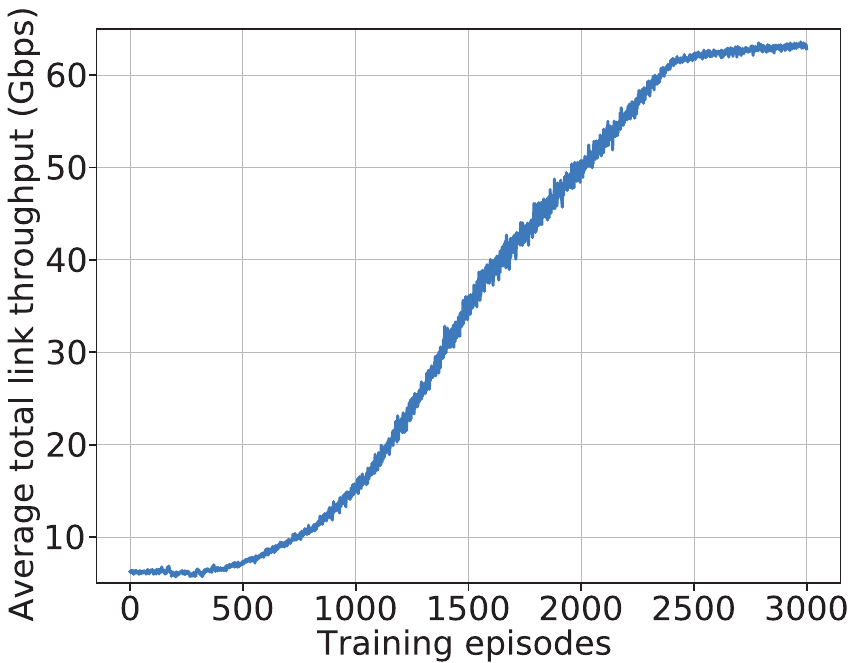}
\centerline{\quad\small (b) $\delta=0.2$}
\end{minipage}
\label{fig:mini:subfig:c}
\begin{minipage}[t]{0.19\linewidth}
\centering
\includegraphics[width=1\columnwidth,height=1.1in]{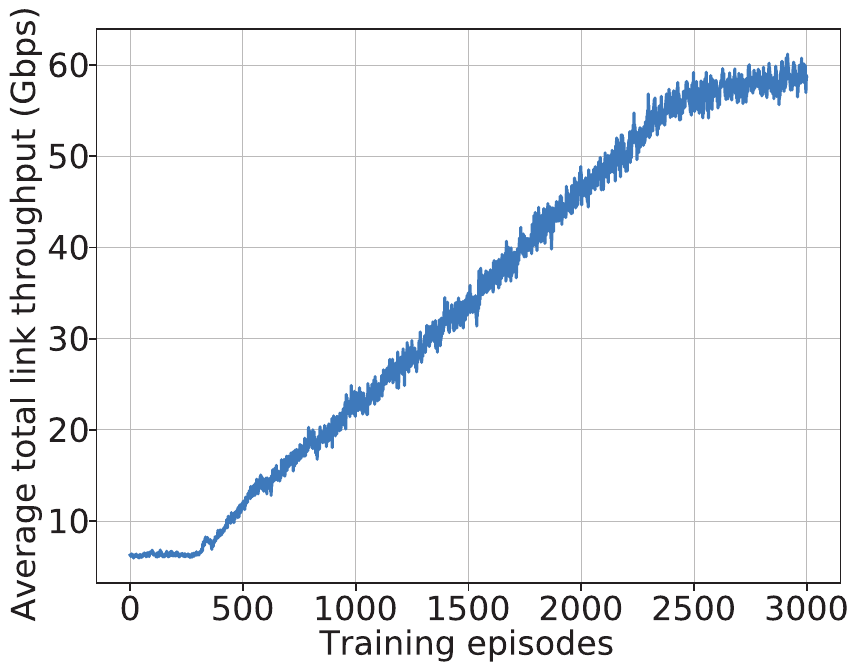}
\centerline{\quad\small (c) $\delta=0.5$}
\end{minipage}
\label{fig:mini:subfig:d}
\centering
\begin{minipage}[t]{0.19\linewidth}
\centering
\includegraphics[width=1\columnwidth,height=1.1in]{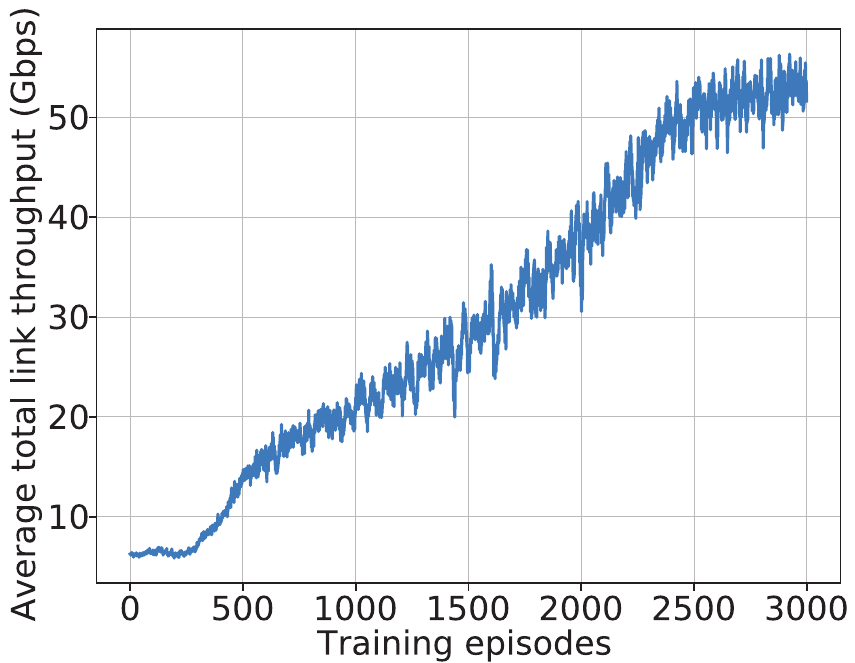}
\centerline{\quad\small (d) $\delta=0.8$}
\end{minipage}%
\label{fig:mini:subfig:e}
\begin{minipage}[t]{0.19\linewidth}
\centering
\includegraphics[width=1\columnwidth,height=1.1in]{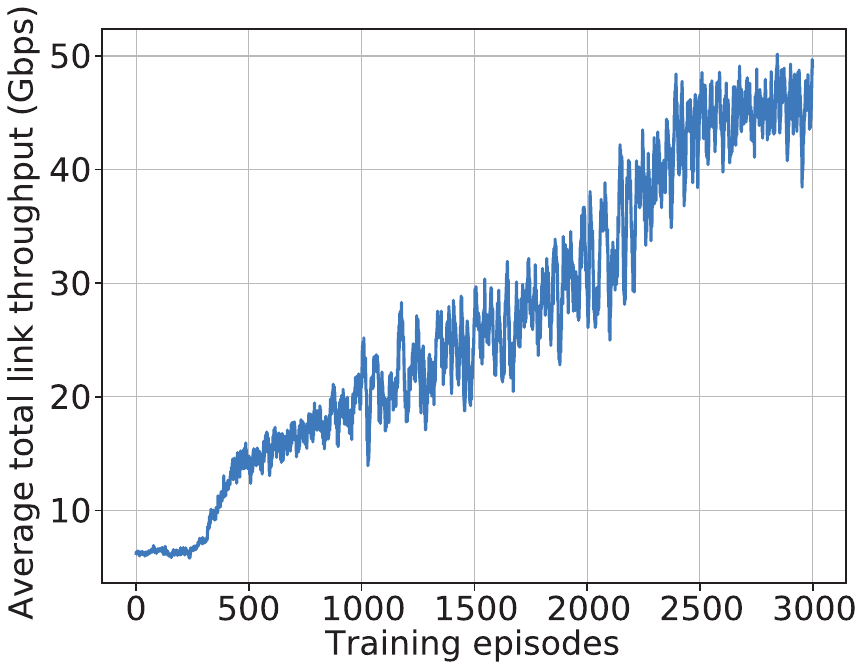}
\centerline{\quad\small (e) $\delta=1$}
\end{minipage}
\caption{Convergence performance on the choice of $\delta$ for MADDQN.}
\vspace*{-5mm}
\label{fig:4}
\end{figure*}

Consider a two-tier HetNet as shown in Fig.~\ref{fig:3}, where the SBSs and UEs are randomly distributed within a radius of 200m centered at the MBS. In the HetNet, small cells are densely deployed, and we set the density of SBSs to be more than 100 BSs/km$^2$ accordingly~\cite{49,50}. MmWave communications are used in the HetNet, where the access links and the backhaul links use the 28 GHz band and the 73 GHz band, respectively. The link parameters are determined by fitting the equations to the measured data in downtown Manhattan via maximum likelihood estimation~\cite{42b}. The settings of simulation parameters are summarized in Table~\ref{Simulation}.

In the MADDQN method, the Q-evaluate network of each UE has the same structure as the Q-target network, i.e., having three fully connected hidden layers with 400, 350, and 300 neurons, respectively. The replay memory size is 150$,$000 and the minibatch size is 1$,$000. Rectified linear unit (ReLU), i.e., $\mbox{ReLU}(x)=\max(x,0)$, is adopted as the activation function. The RMSProp optimizer~\cite{51} is used to update the Q-evaluate network, where the learning rate is set to 0.0001. The discount rate $\gamma$ is set to 0.9. The total training steps $T$ of each episode is 1$,$000, and the total number of episodes $E$ is set on the basis of ensuring the convergence of the algorithm. The exploration rate $\varepsilon$ is set to attenuate linearly from 1 to 0.002 with the increase of episode $e$ over the first $80\%E$ training episodes and be stable at 0.002 afterwards.

It is worth noting that the dynamic changes of the system are reflected in the random access link blockage, which results in dynamic changes in the access link state $c_{ji,t}$ over time. The probability of each access link state is given by~(\ref{eq7}-\ref{eq9}). For the proposed MADDQN method, we first perform algorithm training using Algorithm~\ref{alg.a} and then test the trained MADDQN. In the training stage, the state of the access link between each UE and each SBS changes once at each training time step. Since each episode contains 1$,$000 training steps, the access link states change 1$,$000 times in each episode, which facilitates the MADDQN algorithm to learn the underlying correlation between the dynamics of the link blockage, the joint optimization strategies, and the system throughput performance. In the testing stage, we test the performance of the algorithm over 1$,$000 time steps, and the access link state between each UE and each SBS changes with time steps. At each time step, all UEs make quick decisions with their trained Q-networks based on the current access link state.

To show the advantage of our proposed algorithm in the improvement of link throughput, the proposed user association and backhaul bandwidth allocation scheme based on trained MADDQN is compared with the three baseline schemes for HetNets and one baseline scheme for the macrocell:
\begin{enumerate}
  \item \textbf{HL} (heuristic based user association and load based backhaul bandwidth allocation):
  the user association scheme is proposed in~\cite{52}, which is shown in Algorithm~\ref{alg.b}.
  When the access link between SBS $j$ and UE $i$ is not in the outage state, we treat it as a feasible user association. The algorithm first orders all the feasible user associations according to their respective SNRs, and then validates in order whether the current association can increase the total access link throughput of the HetNet.
  The MBS allocates backhaul bandwidth proportional to the load on each small cell.
\begin{algorithm}[htbp]
\caption{Heuristic scheme for User Association}	
\label{alg.b}
\begin{algorithmic}[1]
\STATE Set $y_{ij,t} = 0, \forall i \in U , \forall j \in B$ at time $t$;
\STATE Get the $SNR_{ij,t}$ if access link state $c_{ji,t} \ne 1$ at time $t$;
\STATE Sort the $SNR_{ij,t}$ values in descending order into $Z_t = \{z_{1,t} ,z_{2,t},...,z_{k,t},...,z_{K,t}\}$. $k = \phi(i,j)$ denotes the mapping between $k$ and the SBS-UE pair $(i,j)$;
\STATE Initialize $R^{ac}_t = 0$;
\WHILE{$k \leq K$}
\STATE Set $y_{k,t} = 1$;
\STATE Compute $R^{ac}_t(k)$;
\IF{$R^{ac}_t(k) > R^{ac}_t(k-1)$ and constraint~(\ref{eq23}) is satisfied }
\STATE Set $y_{k,t} = 1$;
\ELSE
\STATE Set $y_{k,t} = 0$;
\ENDIF
\ENDWHILE
\end{algorithmic}
\end{algorithm}

  \item \textbf{SL} (SNR based user association and load based
  backhaul bandwidth
  allocation): each UE is associated with the SBS which can provide the maximum SNR, and the MBS allocates backhaul bandwidth proportional to the load on each small cell. If the number of UEs requested to associate with a SBS exceeds $N_s$, the SBS selects $N_s$ UEs with the maximum SNR for access, and other UEs select the SBS providing the maximum SNR among the remaining SBSs to associate.

  \item \textbf{DA} (distance based user association and equal
  backhaul bandwidth
  allocation): Each UE is associated with the nearest SBS, and the backhaul bandwidth is evenly allocated to each SBS. If the number of UEs requested to associate with a SBS exceeds $N_s$, the SBS selects the nearest $N_s$ UEs for access, and other UEs select the nearest SBS among the remaining SBSs to associate.

  \item \textbf{MBS-only} (MBS serves UEs directly): There are no small cells deployed in the macro cell. Each UE is associated with the MBS. The MBS communicates with UEs at 28 GHz
   band, and allocates equal bandwidth to each UE.

\end{enumerate}

\vspace{-3mm}
\subsection{Choice of the Selfish Factor }\label{S5-2}

Since the choice of the selfish factor $\delta$ defined in~(\ref{eq32}) has a significant impact on the throughput performance of our scheme, we now investigate it for better choice.

\begin{figure}[t]
\begin{minipage}[t]{0.45\linewidth}
\centering
\includegraphics[width=1.1\columnwidth,height=1.3in]{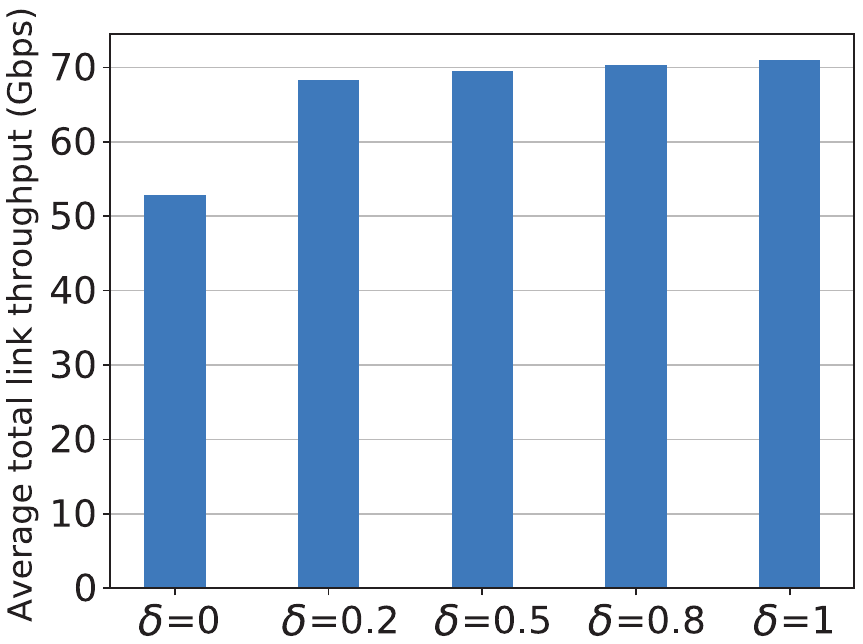}
\centerline{\quad\small (a)}
\end{minipage}%
\hspace{0.2in}
\begin{minipage}[t]{0.45\linewidth}
\centering
\includegraphics[width=1.1\columnwidth,height=1.3in]{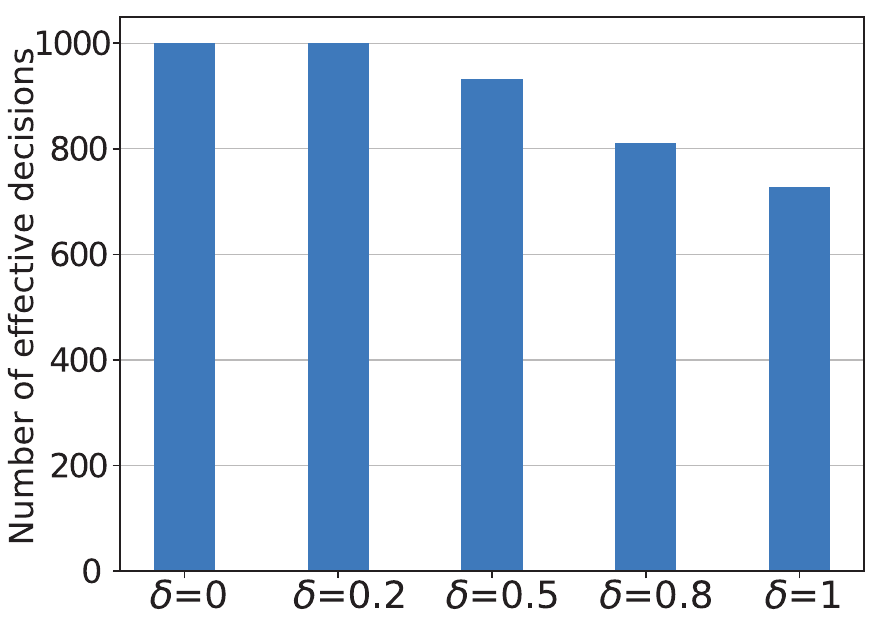}
\centerline{\quad\small (b)}
\end{minipage}
\caption{The (a) average total link throughput of effective decisions and (b) number of effective decisions achieved by MADDQN under different $\delta$.}
\vspace*{-5mm}
\label{fig:5}
\end{figure}

First, in Fig.~\ref{fig:4}, we test the effect of different $\delta$ on the convergence performance of the algorithm. There are 30 UEs, 20 SBSs and a MBS in the HetNet as Fig.~\ref{fig:3} shows, and the number of training episodes $E$ is set to 3$,$000. The average link throughput of each episode represents the average of the link throughput of 1$,$000 training steps in this episode. If the action selection at a training step does not satisfy the constraint~(\ref{eq22}) and (\ref{eq23}), we set the throughput of this training step to 0. As we can see, when $\delta$ is 0 and 0.2, the fluctuation of convergence curve is small, and it can converge stably in about 2400 training episodes. However, as $\delta$ increases, the convergence curve fluctuates more and more sharply. In addition, the convergence value of the average link throughput is maximum at $\delta=0.2$ and decreases with increasing $\delta$ when $\delta>0.2$. To better explain these observations, we test the trained MADDQN of different $\delta$ in the same scenario. The test results are shown in Fig.~\ref{fig:5}.

Fig.~\ref{fig:5} (a) shows the average total link throughput of effective decisions achieved by the trained MADDQN over 1$,$000 time steps under different $\delta$.
The effective decision means that the scheme made by MADDQN can meet the constraints~(\ref{eq22}) and (\ref{eq23}). It can be seen that with the increase of $\delta$, the average total link throughput for effective decisions is improved. This is because as $\delta$ increases, each UE tends to occupy more resource blocks to improve its throughput. This is reflected in the larger $l_{ij,t}$ in action selection. The utilization of backhaul resources in the network is more sufficient at this time, thus improving the average throughput of the network.

However, the increase in $\delta$ also brings a problem. Fig.~\ref{fig:5} (b) shows the number of effective decisions made by MADDQN in 1$,$000 time steps under different $\delta$. As we can see, when $\delta$ is 0 and 0.2, all the 1$,$000 decisions are effective decisions. But when $\delta$ increases to 0.5, 0.8, and 1, the proportion of effective decisions decreases to $93.2\%$, $81.1\%$, and $72.8\%$, respectively. This is because as $\delta$ increases, the UE's aggressive resource block occupancy strategy causes the allocated backhaul resources to exceed the total available backhaul resources in the network. For the actual communication system, this situation will cause network congestion, which is unacceptable. Based on the observations in Fig.~\ref{fig:5}, we can see that setting $\delta$ to 0.2 achieves a balance between improving throughput and satisfying the backhaul resource constraint. Therefore, $\delta=0.2$ is adopted in the subsequent simulations.

\vspace{-3mm}
\subsection{Comparison With Other Schemes }\label{S5-3}

\begin{figure}[htbp]
	\centering
	\includegraphics*[width=2.4in]{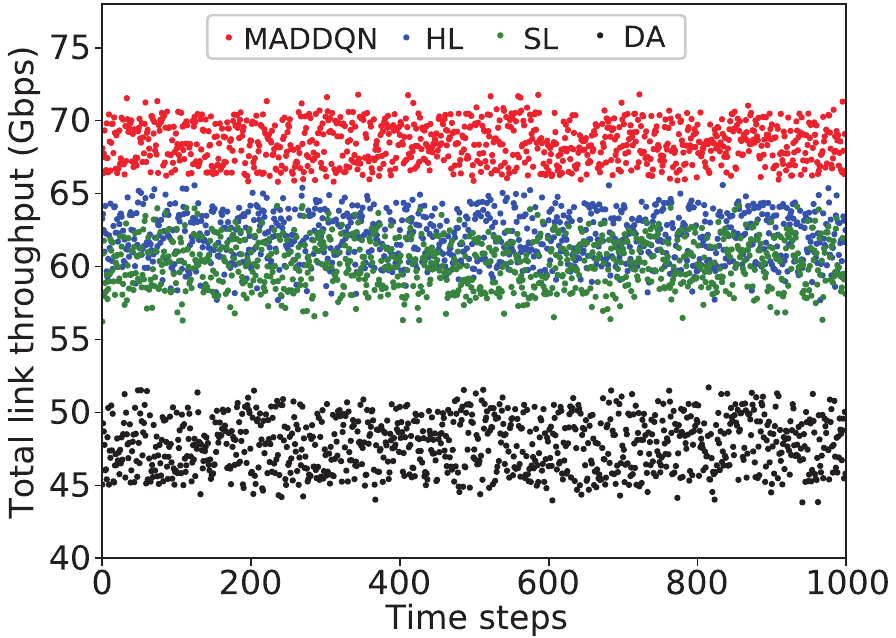}
    \vspace*{-3mm}
	\caption{Total link throughput with different time steps.}
	\label{fig:7}
\vspace*{-3mm}
\end{figure}

Fig.~\ref{fig:7} shows the total link throughput performances of the four schemes at 1$,$000 time steps in the scenario shown in Fig.~\ref{fig:3}. As we can see, since the access link state changes over time steps, the total link throughput of all the scheme varies at different time steps. This intuitively reflects the impact of the dynamic changes in the access link state on the system performance. Besides, in 1$,$000 time steps, MADDQN can always achieve the highest total link throughput compared with the other three baseline schemes. This means that our algorithm can better adapt to the dynamic mmWave scenario. Compared with the other three schemes, the average total link throughput of MADDQN of 1$,$000 time steps achieves improvements of 10.1$\%$, 13.2$\%$, and 42.7$\%$, respectively.

In addition, as shown in Fig.~\ref{fig:7}, the performance of the HL scheme is similar to that of the SL scheme. This is because both HL and SL schemes make user association decisions based on the SNR information of the network, and their backhaul resource allocation schemes are the same. The HL scheme, however, takes into account the impact of the user association strategy on the throughput of the access links, so the throughput performance is slightly better than that of the SL scheme. As for the DA scheme, it cannot adjust the strategies according to the network dynamics, so the performance is the worst.

\begin{figure*}[t]
\begin{minipage}[t]{0.25\linewidth}
\centering
\includegraphics[width=1\columnwidth,height=1.3in]{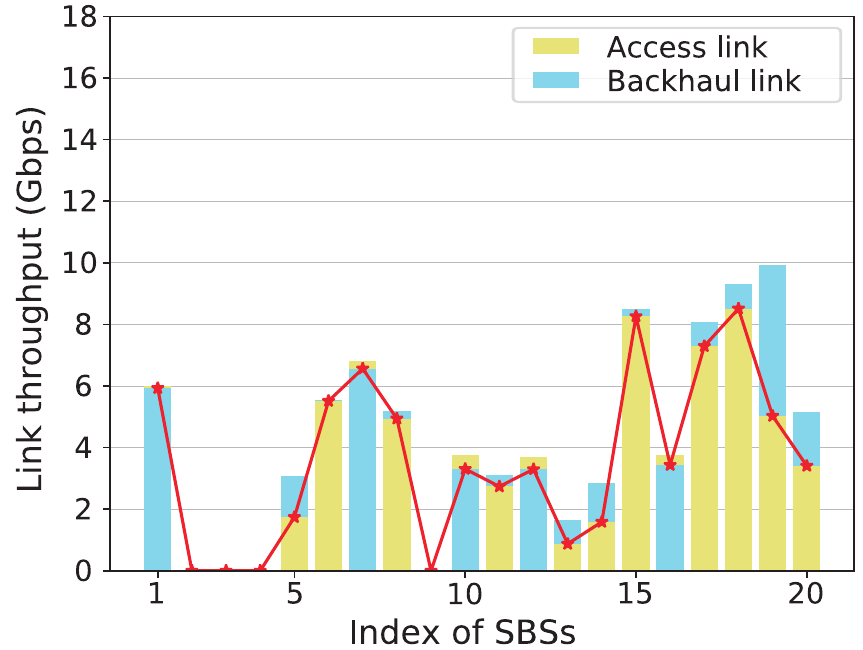}
\centerline{\qquad\small (a)}
\end{minipage}%
\begin{minipage}[t]{0.24\linewidth}
\centering
\includegraphics[width=1\columnwidth,height=1.3in]{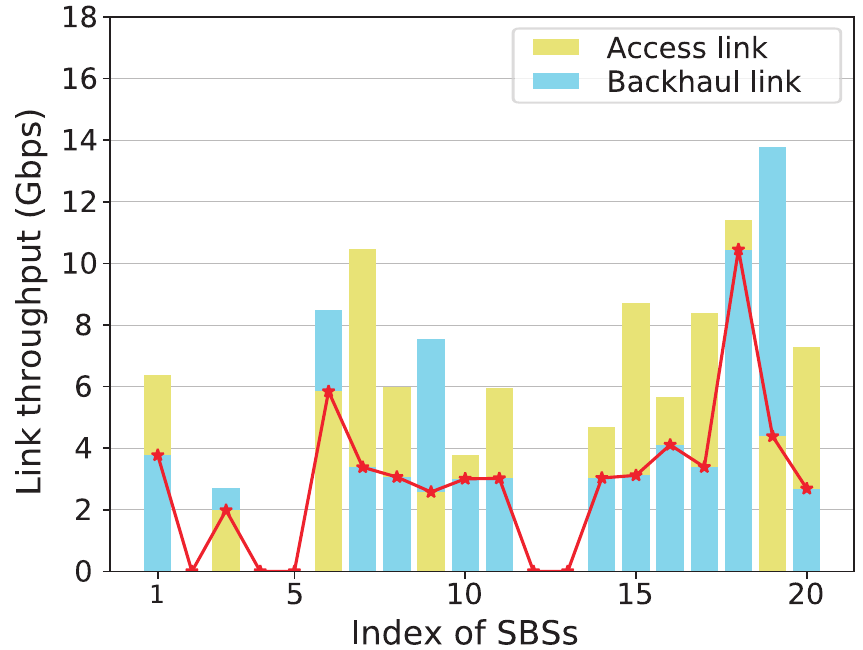}
\centerline{\qquad\small (b)}
\end{minipage}
\begin{minipage}[t]{0.25\linewidth}
\centering
\includegraphics[width=1\columnwidth,height=1.3in]{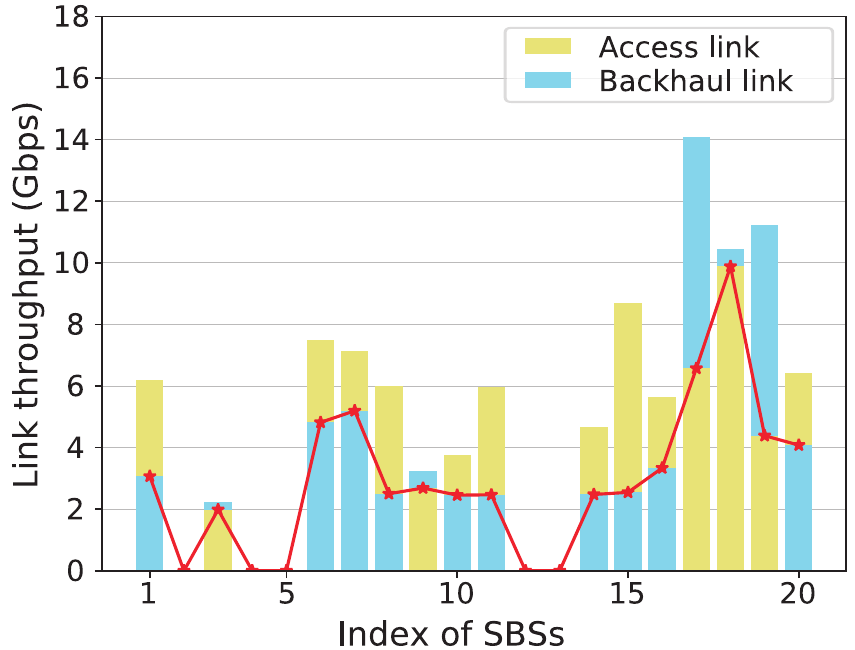}
\centerline{\qquad\small (c)}
\end{minipage}
\begin{minipage}[t]{0.24\linewidth}
\centering
\includegraphics[width=1\columnwidth,height=1.3in]{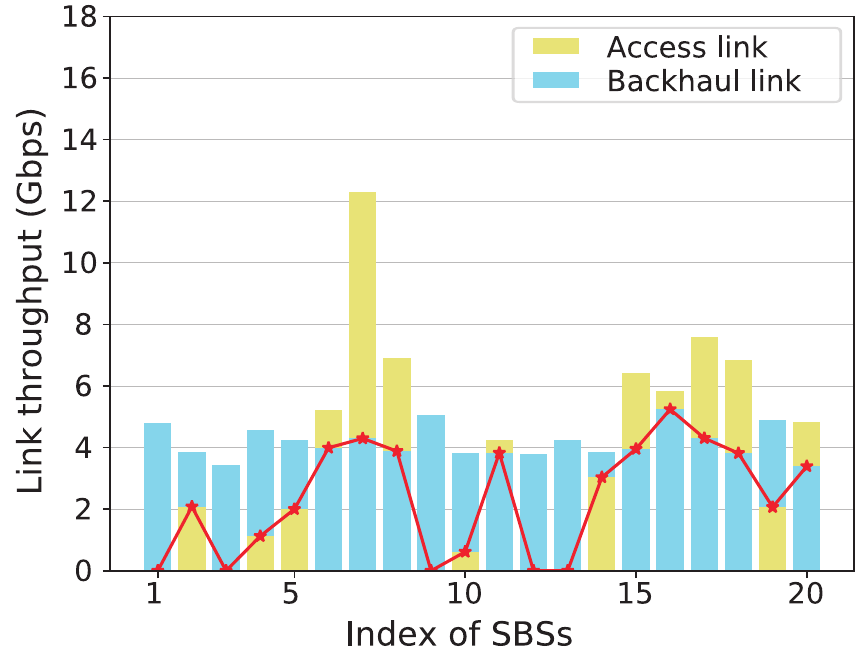}
\centerline{\qquad\small (d)}
\end{minipage}
\caption{Link throughput of each SBS at one time step under the four schemes: (a) MADDQN, (b) HL, (c) SL, and (d) DA.}
\vspace*{-5mm}
\label{fig:8}
\end{figure*}

Then, under different schemes, we test the access and backhaul link throughput of each SBS at one time step as Fig.~\ref{fig:8} shows. We can intuitively observe whether the access link throughput and the backhaul link throughput are matched for each SBS. The red mark in Fig.~\ref{fig:8} represents the actual link throughput of each SBS, determined by the small one between the backhaul link throughput and the access link throughput. From an overall view, MADDQN is most advantageous in balancing the access and backhaul link throughput, which is attributed to the joint design of access and backhaul. As Fig.~\ref{fig:8} (a) shows, SBSs with higher access throughput can always be allocated more backhaul resources to obtain higher backhaul link throughput. Accordingly, higher actual link throughput of these SBSs can be achieved. This explains why MADDQN has the best throughput performance in Fig.~\ref{fig:7}.

In comparison, the SL and HL scheme adopts the load based backhaul resource allocation strategy. However, due to the difference between the access link states of different UEs, the SBS that serves more UEs does not necessarily achieve higher access link throughput, such as the SBS 9 and 19 in Fig.~\ref{fig:8} (b) and the SBS 17 and 19 in Fig.~\ref{fig:8} (c). Allocating more backhaul resources to these SBSs can not increase the actual link throughput. Conversely, it is harmful to the increase of the actual link throughput because of reducing the available backhaul resources for the remaining SBSs. As we can see in Fig.~\ref{fig:8} (b) and (c), the actual link throughput of some SBSs is only half of the access link or backhaul link throughput, which limits the actual link throughput of the HetNet.

Furthermore, the user association and backhaul resource allocation of the DA scheme are independent of each other.
In Fig.~\ref{fig:8} (d),
since SBS 1, 3, 9, 12, and 13 have no UE associated, the actual link throughput of these SBSs is equal to 0.
As the backhaul resources are evenly allocated among the SBSs, the backhaul resources on these SBSs are not utilized.
At the same time,
the evenly allocated backhaul resources can not meet the requirements of the remaining SBSs with high access link throughput, such as SBS 7, 8, 15, 17, and 18.
Therefore, the actual link throughput of these SBSs is limited by the backhaul resources.

\begin{figure}[!t]
	\centering
	\includegraphics*[width=2.2in]{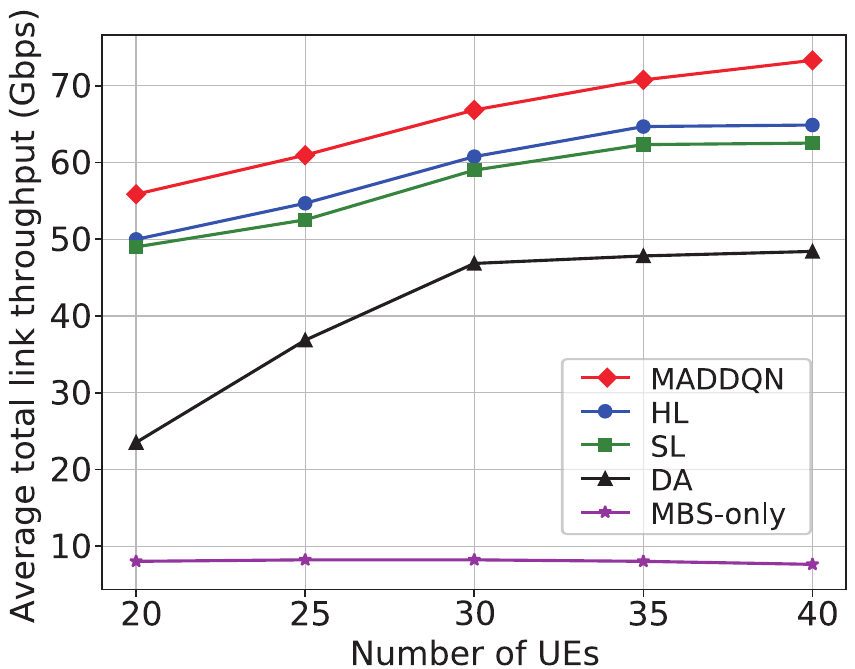}
    \vspace{-3mm}
	\caption{Average total link throughput under different numbers of UEs.}
	\label{fig:10}
\vspace{-3mm}
\end{figure}

In order to test the scalability of our MADDQN scheme, we evaluate the total link throughput performance in different scenarios. First, under different numbers of UEs, the throughput performance of the five schemes is shown in Fig.~\ref{fig:10}. There are 20 SBSs and a MBS in the HetNet. When the number of UEs is 20, 25, 30, 35, and 40, $l_{max}$ is set to 18, 15, 12, 10, and 9, respectively. The number of training episodes $E$ is set to 3$,$000. After the training of MADDQN, we test the average total link throughput over 1$,$000 time steps of five schemes. As shown in Fig.~\ref{fig:10}, the average total link throughput of MADDQN under different numbers of UEs is always higher than that of the other four baseline schemes. When the number of UEs is 20, 25, 30, 35 and 40, the improvements in the total link throughput of MADDQN over HL are 11.8$\%$, 11.5$\%$, 10.1$\%$, 9.39$\%$, and 12.99$\%$, respectively.

Besides, we can see that with the increasing number of UEs, the average total link throughput of the four schemes for HetNets increases. However, due to the limited resources of the HetNet, the improvements of the four schemes get smaller when more UEs are served by the HetNet. In contrast with the other three schemes, the network throughput growth of the MADDQN scheme is less constrained. The reason is that the MADDQN scheme optimizes the user association and backhaul bandwidth allocation jointly, which allows for more flexible and efficient utilization of the resource in the HetNet. In addition, we can observe that the dense deployment of mmWave small cells can provide a significant boost in system throughput compared to the MBS-only scheme. For example, when the number of UEs is 30, the average total link throughput of MADDQN scheme for HetNet is 7.88 times higher than that of the MBS-only scheme. This is because with the dense deployment of small cells, the distance from UEs to SBSs can be reduced, which achieves higher SNR and enhances network coverage. Besides, UEs in different small cells can be served in the same spectrum through frequency reuse, so the network spectrum efficiency can be improved.

\begin{figure}[!t]
\centering
\includegraphics*[width=2.2in]{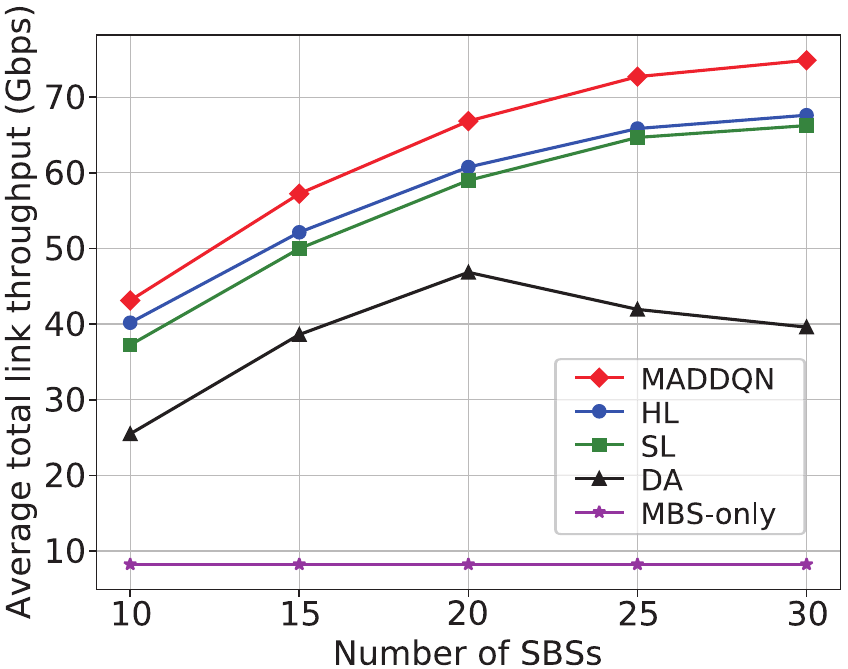}
\vspace{-3mm}
\caption{Average total link throughput under different number of SBSs.}
\label{fig:11}
\vspace{-5mm}
\end{figure}

Finally, we examine the performance of the four schemes for HetNets under different numbers of SBSs, and the results are presented in Fig.~\ref{fig:11}. The MBS-only scheme is also used as a comparison. There are 30 UEs and a MBS in the HetNet. The number of SBSs is set to 10, 15, 20, 25, and 30, respectively and the number of training episodes $E$ is 3$,$000. Similarly, performance of five schemes is evaluated with the average total link throughput of 1$,$000 time steps. As shown in Fig.~\ref{fig:11}, the average total link throughput of MADDQN under different numbers of SBSs is higher than that of the other four baseline schemes. When the number of SBSs is 10, 15, 20, 25, and 30, the improvements in average total link throughput of MADDQN over HL are 7.3$\%$, 9.7$\%$, 10.1$\%$, 10.4$\%$ and 10.7$\%$, respectively.

In addition, under different number of SBSs in the HetNet, the throughput of the four schemes for HetNets is always significantly greater than that of the MBS-only scheme. In the HetNet, with the increasing number of SBSs, there are more available access bandwidth resources for UEs due to the frequency reuse in different small cells.
Besides, UEs will find SBSs closer to them with better service quality. The access load of each SBS is also reduced. So as we can see in Fig.~\ref{fig:11}, the throughput of the MADDQN, HL and SL schemes can be improved. Compared with the MBS-only scheme, higher throughput gains can be achieved with these three schemes when there are more SBSs in the HetNet. However, since the increase in the number of small cells causes more inter-cell interference and the backhaul resources in the network are limited, the improvement of the total link throughput gradually slows down with the increase in the number of SBS.

Unlike the above three schemes, the average total link throughput of the DA scheme decreases with the increasing number of SBSs when there are more than 20 SBSs in the HetNet. The reason is that when there are more SBSs in the HetNet, the available backhaul resources for each SBS become less due to the equal backhaul bandwidth allocation. Besides, there may be more SBSs with no UE association in the HetNet, which causes more serious waste of backhaul resources. Although the access link throughput has been improved by the dense deployment of SBSs, the actual total link throughput is limited by the backhaul link throughput.

In contrast with the other four schemes, the MADDQN scheme we proposed guarantees the balance of the access and backhaul throughput with the joint design of access and backhaul, thus providing greater performance gains than the other four schemes when more SBSs are deployed in the HetNet. This also reflects that MADDQN is more suitable for the mmWave small cells dense deployment scenarios.

\vspace{-2mm}
\section{Conclusions}\label{S6}

In this paper, we investigated the problem of user association and backhaul bandwidth resource allocation in two-tier mmWave HetNets, where small cells are densely deployed and two different mmWave bands are allocated to access and backhaul. We formulated the joint user association and backhaul resource allocation problem and then transformed the problem into a Markov game. A joint design scheme based on MADRL was proposed for the maximization of the long-term total link throughput. The proposed scheme treated each UE as an agent and allowed each UE to learn the optimal policy autonomously by DDQN based on its state observations. Through extensive training, each UE can dynamically adjust the policy to the time-varying link state. Simulation results showed that the proposed MADRL scheme could adapt to the dynamic mmWave link states, and achieve high total link throughput under various system configurations, and outperformed four baseline schemes.

Due to the appropriate deployment of SBSs, we assume the LOS transmission between MBS and SBSs in this paper. However, the mmWave backhaul link may also be affected by link blockage in real communication systems. Therefore, we will consider a more realistic channel model for backhaul in future work. Besides, multi-hop wireless backhaul helps to further improve network coverage and combat mmWave link blockage, enabling more flexible backhaul connectivity. In the IAB networks with multi-hop backhaul, the backhaul path selection is introduced and the end-to-end latency, throughput, and fairness are critical performance metrics. The joint design of access and backhaul in such networks is an open research problem. Moreover, we will evaluate the impact of introducing direct communications between MBS and UEs on the system performance and design an effective joint optimization algorithm for the system. In addition, we will include UE mobility and changes in the number of UEs as additional cases of environment dynamics for future work.

\vspace{-2mm}


\bibliographystyle{IEEEtran}

\begin{thebibliography}{}
\providecommand{\url}[1]{#1}
\csname url@samestyle\endcsname
\providecommand{\newblock}{\relax}
\providecommand{\bibinfo}[2]{#2}
\providecommand{\BIBentrySTDinterwordspacing}{\spaceskip=0pt\relax}
\providecommand{\BIBentryALTinterwordstretchfactor}{4}
\providecommand{\BIBentryALTinterwordspacing}{\spaceskip=\fontdimen2\font plus
\BIBentryALTinterwordstretchfactor\fontdimen3\font minus
  \fontdimen4\font\relax}
\providecommand{\BIBforeignlanguage}[2]{{%
\expandafter\ifx\csname l@#1\endcsname\relax
\typeout{** WARNING: IEEEtran.bst: No hyphenation pattern has been}%
\typeout{** loaded for the language `#1'. Using the pattern for}%
\typeout{** the default language instead.}%
\else
\language=\csname l@#1\endcsname
\fi
#2}}
\providecommand{\BIBdecl}{\relax}
\BIBdecl

\end{thebibliography}


\begin{thebibliography}{10}
{
\bibliographystyle{IEEEtran}

\bibitem{01}
M.N. Islam, A. Sampath, A. Maharshi, O. Koymen and N.B. Mandayam, ``Wireless backhaul node placement for small cell networks,'' in \emph{Proc. IEEE 48th Annu. CISS}, Princeton, NJ, Mar. 2014, pp. 1--6.

\bibitem{02}
Y. Niu, Y. Li, D. Jin, L. Su, and A.V. Vasilakos, ``A survey of millimeter wave communications (mmWave) for 5G: opportunities and challenges,'' \emph{Wireless Netw.}, vol.~21, no.~8,  pp. 2657--2676, Apr. 2015.

\bibitem{03}
R. Baldemair \emph{et al.}, ``Ultra-dense networks in millimeter-wave frequencies,'' \emph{IEEE Commun. Mag.}, vol.~53, no.~1, pp. 202--208, Jan. 2015.

\bibitem{04}
D. L\'{o}pez-P\'{e}rez, M. Ding, H. Claussen and A. H. Jafari, ``Towards 1 Gbps/UE in Cellular Systems: Understanding Ultra-Dense Small Cell Deployments,'' \emph{IEEE Commun. Surv. Tutor.}, vol.~17, no.~4, pp. 2078--2101, Fourthquarter 2015.

\bibitem{05}
M. Polese \emph{et al.}, ``Integrated Access and Backhaul in 5G mmWave Networks: Potential and Challenges,'' \emph{IEEE Commun. Mag.}, vol.~58, no.~3, pp. 62--68, Mar. 2020.

\bibitem{06}
3GPP, ``Study on Integrated Access and Backhaul,'' TR 38.874, 2018.

\bibitem{07b}
C. Saha, M. Afshang and H. S. Dhillon, ``Bandwidth Partitioning and Downlink Analysis in Millimeter Wave
Integrated Access and Backhaul for 5G,'' \emph{IEEE Trans. Wireless Commun.}, vol.~17, no.~12, pp. 8195--8210, Dec. 2018.

\bibitem{08}
N. Wang, E. Hossain, and V.K. Bhargava, ``Joint downlink cell association and bandwidth allocation for wireless backhauling in two-tier HetNets with large-scale antenna arrays,'' \emph{IEEE Trans. Wireless Commun.}, vol.~15, no.~5, pp. 3251--3268, May 2016.

\bibitem{08b}
R. Liu, Q. Chen, G. Yu and G. Y. Li, ``Joint User Association and Resource Allocation for Multi-Band Millimeter-Wave Heterogeneous Networks,'' \emph{IEEE Trans. Commun.}, vol.~67, no.~12, pp. 8502--8516, Dec. 2019.

\bibitem{09}
Y. Niu, C. Gao, Y. Li, L. Su, D. Jin and A. V. Vasilakos, ``Exploiting Device-to-Device Communications in Joint Scheduling of Access and Backhaul for mmWave Small Cells,'' \emph{IEEE J. Sel. Areas Commun.}, vol.~33, no.~10, pp. 2052--2069, Oct. 2015.

\bibitem{10}
Y. Liu, L. Lu, G. Y. Li, Q. Cui and W. Han, ``Joint User Association and Spectrum Allocation for Small Cell Networks With Wireless Backhauls,'' \emph{IEEE Wireless Commun. Lett.}, vol.~5, no.~5, pp. 496--499, Oct. 2016.

\bibitem{11}
A. Khodmi, S. B. Rejeb, N. Agoulmine and Z. Choukair, ``A Joint Power Allocation and User Association Based on Non-Cooperative Game Theory in an Heterogeneous Ultra-Dense Network,'' \emph{IEEE Access}, vol.~7, pp. 111790--111800, Aug. 2019.

\bibitem{12}
Z. Su \emph{et al.}, ``User association and wireless backhaul bandwidth allocation for 5G heterogeneous networks in the millimeter-wave band,'' \emph{China Commun.}, vol.~15, no.~4, pp. 1-13, Apr. 2018.

\bibitem{13}
Y. Liu, X. Fang, P. Zhou, and K. Cheng, ``Coalition game for user association and bandwidth allocation in ultra-dense mmWave networks,'' in \emph{Proc. IEEE/CIC ICCC}, Chengdu, China, Dec. 2017, pp. 1--5.

\bibitem{13b}
N. C. Luong \emph{et al.}, ``Applications of Deep Reinforcement Learning in Communications and Networking: A Survey,'' \emph{IEEE Commun. Surv. Tutor.}, vol.~21, no.~4, pp. 3133--3174, Fourthquarter 2019.

\bibitem{14}
X. Shen, J. Gao, W. Wu, M. Li, C. Zhou and W. Zhuang, ``Holistic Network Virtualization and Pervasive Network Intelligence for 6G,'' \emph{IEEE Commun. Surv. Tutor.}, vol.~24, no.~1, pp. 1--30, Firstquarter 2022.

\bibitem{15}
X. Shen \emph{et al.}, ``AI-Assisted Network-Slicing Based Next-Generation Wireless Networks,'' \emph{IEEE Open J. Veh. Technol.}, vol.~1, pp. 45--66, Jan. 2020.

\bibitem{16}
K. Xiao, S. Mao, and J.K. Tugnait, ``TCP-Drinc: Smart congestion control based on deep reinforcement learning,'' {\em IEEE Access}, vol.~7, no.~1, pp. 11892--11904, Jan. 2019.

\bibitem{17}
S. Shen, T. Zhang, S. Mao, and G.-K. Chang, ``DRL-based channel and latency aware radio resource allocation for 5G service-oriented RoF-mmWave RAN,'' {\em J. Light. Technol.}, vol.~39, no.~18, pp. 5706--5714, Sept. 2021.

\bibitem{20}
R.S. Sutton and A.G. Barto, \emph{Reinforcement Learning: An Introduction}, Cambridge, MA: MIT press, 2018.

\bibitem{21}
C.J.C.H. Watkins and P. Dayan, ``Q-learning,'' \emph{Mach. Learn.}, vol.~5, no.~3/4, pp. 279--292, May 1992.

\bibitem{22}
V. Mnih, \emph{et al.}, ``Human-level control through deep reinforcement learning.'' \emph{Nature}, vol.~518, no.~7540, pp. 529--533, Feb. 2015.

\bibitem{23}
G. Tesauro, ``Extending Q-learning to general adaptive multi-agent systems.'' in \emph{Proc. NIPS}, Vancouver and Whistler, Canada, Dec. 2003, pp. 871--878.

\bibitem{26}
A. Alwarafy, M. Abdallah, B. S. \c{C}iftler, A. Al-Fuqaha and M. Hamdi, ``The Frontiers of Deep Reinforcement Learning for Resource Management in Future Wireless HetNets: Techniques, Challenges, and Research Directions,'' \emph{IEEE Open J. Commun. Soc.}, vol.~3, pp. 322--365, Feb. 2022.

\bibitem{27}
D. Fooladivanda and C. Rosenberg, ``Joint Resource Allocation and User Association for Heterogeneous Wireless Cellular Networks,'' \emph{IEEE Trans. Wireless Commun.}, vol.~12, no.~1, pp. 248--257, Jan. 2013.

\bibitem{28}
Y. Lin, W. Bao, W. Yu and B. Liang, ``Optimizing User Association and Spectrum Allocation in HetNets: A Utility Perspective,'' \emph{IEEE J. Sel. Areas Commun.}, vol.~33, no.~6, pp. 1025--1039, June 2015.

\bibitem{29}
Y. Chen, J. Li, W. Chen, Z. Lin and B. Vucetic, ``Joint User Association and Resource Allocation in the Downlink of Heterogeneous Networks,'' \emph{IEEE Trans. Veh. Technol.}, vol.~65, no.~7, pp. 5701--5706, July 2016.

\bibitem{30}
B. Zhuang, D. Guo and M. L. Honig, ``Energy-Efficient Cell Activation, User Association, and Spectrum Allocation in Heterogeneous Networks,'' \emph{IEEE J. Sel. Areas Commun.}, vol.~34, no.~4, pp. 823--831, Apr. 2016.

\bibitem{31}
C. Chaieb, Z. Mlika, F. Abdelkefi and W. Ajib, ``On the user association and resource allocation in hetnets with mmWave base stations,'' in \emph{Proc. IEEE PIMRC}, Montreal, QC, Canada, Oct. 2017, pp. 1--5.

\bibitem{32}
K. Khawam, S. Lahoud, M. E. Helou, S. Martin and F. Gang, ``Coordinated Framework for Spectrum Allocation and User Association in 5G HetNets With mmWave,'' \emph{IEEE Trans. Mob. Comput.}, vol.~21, no.~4, pp. 1226--1243, Apr. 2022.

\bibitem{34}
W. Hao, M. Zeng, Z. Chu, S. Yang and G. Sun, ``Energy-Efficient Resource Allocation for mmWave Massive MIMO HetNets With Wireless Backhaul,'' \emph{IEEE Access}, vol.~6, pp. 2457--2471, Dec. 2017.

\bibitem{35}
S. Ni, J. Zhao, H. H. Yang and Y. Gong, ``Enhancing Downlink Transmission in MIMO HetNet With Wireless Backhaul,'' \emph{IEEE Trans. Veh. Technol.}, vol.~68, no.~7, pp. 6817--6832, July 2019.

\bibitem{36}
S. Aboagye, A. Ibrahim and T. M. N. Ngatched, ``Frameworks for Energy Efficiency Maximization in HetNets With Millimeter Wave Backhaul Links,'' \emph{IEEE Trans. Green Commun. Netw.}, vol.~4, no.~1, pp. 83--94, Mar. 2020.

\bibitem{37}
M. Feng and S. Mao, ``Dealing with limited backhaul capacity in millimeter-wave systems: A deep reinforcement learning approach,'' \emph{IEEE Commun. Mag.}, vol.~57, no.~3, pp. 50--55, Mar. 2019.

\bibitem{38}
Y. Wei, F. R. Yu, M. Song and Z. Han, ``User Scheduling and Resource Allocation in HetNets With Hybrid Energy Supply: An Actor-Critic Reinforcement Learning Approach,'' \emph{IEEE Trans. Wireless Commun.}, vol.~17, no.~1, pp. 680--692, Jan. 2018.

\bibitem{39}
N. Zhao, Y. -C. Liang, D. Niyato, Y. Pei, M. Wu and Y. Jiang, ``Deep Reinforcement Learning for User Association and Resource Allocation in Heterogeneous Cellular Networks,'' \emph{IEEE Trans. Wireless Commun.}, vol.~18, no.~11, pp. 5141--5152, Nov. 2019.

\bibitem{40}
H. Yang, J. Zhao, K. -Y. Lam, Z. Xiong, Q. Wu and L. Xiao, ``Distributed Deep Reinforcement Learning-Based Spectrum and Power Allocation for Heterogeneous Networks,'' \emph{IEEE Trans. Wireless Commun.}, vol.~21, no.~9, pp. 6935--6948, Sept. 2022.

\bibitem{41}
M. Sana, A. De Domenico, W. Yu, Y. Lostanlen and E. Calvanese Strinati, ``Multi-Agent Reinforcement Learning for Adaptive User Association in Dynamic mmWave Networks,'' \emph{IEEE Trans. Wireless Commun.}, vol.~19, no.~10, pp. 6520--6534, Oct. 2020.

\bibitem{42}
Y. Niu \emph{et al.}, ``Energy-efficient scheduling for mmWave backhauling of small cells in heterogeneous cellular networks,'' \emph{IEEE Trans. Veh. Technol.}, vol.~66, no.~3, pp. 2674--2687, Mar. 2017.

\bibitem{42b}
T. S. Rappaport, G. R. MacCartney, M. K. Samimi and S. Sun, ``Wideband Millimeter-Wave Propagation Measurements and Channel Models for Future Wireless Communication System Design,'' \emph{IEEE Trans. Commun.}, vol.~63, no.~9, pp. 3029--3056, Sept. 2015.

\bibitem{43}
T. Bai and R. W. Heath, ``Coverage and Rate Analysis for Millimeter-Wave Cellular Networks,'' \emph{IEEE Trans. Wireless Commun.}, vol.~14, no.~2, pp. 1100--1114, Feb. 2015.

\bibitem{45}
M. R. Akdeniz \emph{et al.}, ``Millimeter Wave Channel Modeling and Cellular Capacity Evaluation,'' \emph{IEEE J. Sel. Areas Commun.}, vol.~32, no.~6, pp. 1164--1179, June 2014.

\bibitem{45b}
Ajani, T. S., Imoize, A. L. and Atayero, A. A., ``An overview of machine learning within embedded and mobile devices optimizations and applications,'' \emph{Sensors}, vol.~21, no.~13, pp. 4412, June 2021.

\bibitem{45c}
Y. Deng, ``Deep learning on mobile devices: A review,'' in \emph{Proc. SPIE Mobile Multimedia/Image Process., Secur., Appl.}, Oct. 2019, pp. 52--66.

\bibitem{45d}
J. Lee et al., ``On-device neural net inference with mobile GPUs,''  \emph{ arXiv preprint arXiv:1907.01989}, 2019. [Online]. Available: http://arxiv.org/abs/1907.01989


\bibitem{46}
J. Foerster, \emph{et al.}, ``Stabilising experience replay for deep multi-agent reinforcement learning,'' in \emph{Proc. ICML}, Sydney, Australia, Aug. 2017, pp. 1146--1155.

\bibitem{47}
L. Bu\c{s}oniu, R. Babu\v{s}ka, and B. De Schutter. ``Multi-agent reinforcement learning: An overview,'' in \emph{Innovations in Multi-Agent Systems and Applications-1}, D. Srinivasan and L.C. Jain (eds.), vol. 310, pp. 183--221, Berlin, Heidelberg: Springer, 2010.

\bibitem{48}
H. van Hasselt, A. Guez, and D. Silver. ``Deep reinforcement learning with double Q-learning,'' in \emph{Proc. AAAI}, Phoenix, AZ, Feb. 2016, pp. 2094--2100.

\bibitem{49}
T. Zhang, J. Zhao, L. An and D. Liu, ``Energy Efficiency of Base Station Deployment in Ultra Dense HetNets: A Stochastic Geometry Analysis,'' \emph{IEEE Wireless Commun. Lett.}, vol.~5, no.~2, pp. 184--187, Apr. 2016.

\bibitem{50}
T. Ding, M. Ding, G. Mao, Z. Lin, A. Y. Zomaya and D. L\'{o}pez-P\'{e}rez, ``Performance Analysis of Dense Small Cell Networks With Dynamic TDD,'' \emph{IEEE Trans. Veh. Technol.}, vol.~67, no.~10, pp. 9816--9830, Oct. 2018.


\bibitem{51}
S. Ruder, ``An overview of gradient descent optimization algorithms.'' {\em arXiv preprint arXiv:1609.04747}, Sept. 2016. [online] Available: https://arxiv.org/abs/1609.04747.

\bibitem{52}
P. Zhou, X. Fang, X. Wang, Y. Long, R. He and X. Han, ``Deep Learning-Based Beam Management and Interference Coordination in Dense mmWave Networks,'' \emph{IEEE Trans. Veh. Technol.}, vol.~68, no.~1, pp. 592--603, Jan. 2019.
}
\end{thebibliography}

\end{document}